\documentclass[usenatbib]{mnras}

\usepackage{newtxtext,newtxmath}

\usepackage{graphicx}
\usepackage[space]{grffile}
\usepackage{latexsym}
\usepackage{textcomp}
\usepackage{longtable}
\usepackage{multirow,booktabs}
\usepackage{amsfonts,amsmath,amssymb}
\usepackage{url}
\usepackage{hyperref}
\hypersetup{colorlinks=false,pdfborder={0 0 0}}
\newif\iflatexml\latexmlfalse
\usepackage[utf8]{inputenc}
\usepackage[english]{babel}

\usepackage{natbib}

\defcitealias{pmo+14}{P14}

\newcommand{\nqso}{153}  
\newcommand{\lstar}{1.46 \pm 0.11} 
\newcommand{\aval}{1.70 \pm 0.22}
\newcommand{\bestgamma}{1.59 \pm 0.06}  
\newcommand{\bestgtwo}{1.77 \pm 0.06}  
\newcommand{\redchi}{5.7}
\newcommand{\lzfull}{2.6}
\newcommand{\slzfull}{0.4}

\newcommand{\kms}{km s$^{-1}$}
\newcommand{\kmsMpc}{km\,s$^{-1}$\,Mpc$^{-1}$}
\newcommand{\cmm}{cm$^{-2}$}

\newcommand{\Lya}{Ly$\alpha$}
\newcommand{\mLya}{\ensuremath{\textrm{Ly}\alpha}}

\newcommand{\tauLL}{$\tau_{\rm LL}$}
\newcommand{\mtauLL}{\tau_{\rm LL}}
\newcommand{\mloz}{\ell(z)}
\newcommand{\loz}{$\mloz$}
\newcommand{\mlox}{\ell(X)}
\newcommand{\lox}{$\mlox$}
\newcommand{\mrawloz}{\ell(z)_{\rm raw}}
\newcommand{\rawloz}{$\mrawloz$}
\newcommand{\mrawlox}{\ell(X)_{\rm raw}}
\newcommand{\rawlox}{$\mrawlox$}
\newcommand{\mgoz}{g(z)}
\newcommand{\goz}{$\mgoz$}

\newcommand{\mzem}{z_{\rm em}}
\newcommand{\zem}{$\mzem$}
\newcommand{\mzlls}{z_{\rm LLS}}
\newcommand{\zlls}{$\mzlls$}
\newcommand{\mzmin}{z_{\rm min}}
\newcommand{\zmin}{$\mzmin$}
\newcommand{\mzmax}{z_{\rm max}}
\newcommand{\zmax}{$\mzmax$}

\newcommand{\HI}   {{\rm H}\,\textsc{i}}
\newcommand{\HII}  {{\rm H}\,\textsc{ii}}
\newcommand{\HeI}  {{\rm He}\,\textsc{i}}
\newcommand{\HeII} {{\rm He}\,\textsc{ii}}
\newcommand{\HeIII}{{\rm He}\,\textsc{iii}}

\newcommand{\NV}   {{\rm N}\,\textsc{v}}

\newcommand{\OVI}  {{\rm O}\,\textsc{vi}}

\newcommand{\CIV}  {{\rm C}\,\textsc{iv}}

\newcommand{\lya}{Ly$\alpha$}

\newcommand{\NHI}{$N_{\rm HI}$}
\newcommand{\mNHI}{N_{\rm HI}}

\usepackage[usenames,dvipsnames]{color}

\def \mfnhi {f(\mNHI,X)}
\def \fnhi {$\mfnhi$}
\def \mdlogf {d \log \mfnhi/ d \log \mNHI}
\def \dlogf {$\mdlogf$}
\def \mON {O_{\rm HI}(\mNHI, X)}
\def \ON {$\mON$}

\def \llls {$\ell_{\rm{LLS}}(X)$}
\def \mtlya {\tau_{\rm eff}^{\rm Ly\alpha}}
\def \tlya {$\mtlya$}

\def \lmfp {$\lambda_{\rm mfp}^{912}$}
\def \ldla {$\ell_{\rm{DLA}}(X)$}

\newcommand{\lNHI}{\ensuremath{\log(N_\textsc{h\scriptsize{\,i}}/\textrm{cm}^{-2})}}

\title[Lyman limit systems at $\mathbf{z\sim5}$]{
Imprints of the first billion years:  Lyman limit systems at 
$z \sim 5$}

\author[]{Neil H. M. Crighton,$^{1}$ J. Xavier Prochaska ,$^{2}$ Michael T. Murphy$^{1}$, John M. O'Meara$^{3}$,\newauthor
G\'abor Worseck$^{4,5}$, Britton D. Smith$^{6}$\\
$^{1}$Centre for Astrophysics and Supercomputing, Swinburne University of Technology, Hawthorn, Victoria 3122, Australia\\
$^{2}$Astronomy and Astrophysics, UC Santa Cruz, 1156 High St., Santa Cruz, CA 95064, USA\\
$^{3}$Department of Chemistry and Physics, Saint Michael’s College, One Winooski Park, Colchester, VT 05439, USA\\
$^{4}$Max-Planck-Institut f\"ur Astronomie, K\"onigstuhl 17, D-69117 Heidelberg, Germany\\
$^{5}$Institut f\"ur Physik und Astronomie, Universit\"at Potsdam, Karl-Liebknecht-Str.\ 24/25, D-14476 Potsdam, Germany\\
$^{6}$San Diego Supercomputer Center, University of California, San
Diego, 10100 Hopkins Drive, La Jolla, CA 92093\\
}

\pubyear{2018}

\begin{document}
\label{firstpage}
\pagerange{\pageref{firstpage}--\pageref{lastpage}}
\maketitle

\begin{abstract}
Lyman Limit systems (LLSs) trace the low-density circumgalactic medium and the most dense regions of the intergalactic medium, so their number density and evolution at high redshift, just after reionisation, are important to constrain. We present a survey for LLSs at high redshifts, $\mzlls =3.5$--5.4, in the homogeneous dataset of \nqso\ optical quasar spectra at $z \sim 5$ from the Giant Gemini GMOS survey. Our analysis includes detailed investigation of survey biases using mock spectra which provide important corrections to the raw measurements. 
We estimate the incidence of LLSs per unit redshift 
at $z \approx 4.4$ to be $\mloz = \lzfull \pm \slzfull$. 
Combining our results with previous surveys at $\mzlls <4$, the best-fit power-law evolution is $\mloz = \ell_* [(1+z)/4]^\alpha$ with $\ell_* = \lstar$ and $\alpha = \aval$ (68\% confidence intervals). Despite hints in previous $\mzlls <4$ results, there is no indication for a deviation from this single power-law soon after reionization. Finally, we integrate our new results with previous surveys of the intergalactic and circumgalactic media to constrain the hydrogen column density distribution function, \fnhi, over 10 orders of magnitude.  The data at $z \sim 5$ are not well described by the \fnhi\ model previously reported for $z \sim 2$--3 (after re-scaling) and a 7-pivot model fitting the full $z \sim 2$--5 dataset is statistically unacceptable. We conclude that there is significant evolution in the shape of \fnhi\ over this $\sim$2 billion year period.
\end{abstract}

\begin{keywords}
quasars: absorption lines -- cosmological parameters -- cosmology: observations
\end{keywords}

\section{Introduction}

One of the three main classes of \HI\ absorption systems 
seen along quasar sightlines are the Lyman limit systems (LLSs), caused by hydrogen clouds 
with integrated \HI\ column densities (\NHI) large enough to produce optically thick absorption at the Lyman continuum. These clouds are generally highly ionized, with neutral fractions of a few per cent or less \citep[e.g.][]{steidel90,pro99,fumagalli+16a}. 
Their high \HI\ column densities, however, imply an origin in
dense environments such as galaxy haloes, tracing accreting, outflowing, 
and stripped gas, or in the most dense regions of the intergalactic medium (IGM). 

The clear observational signature of an LLS
is a break at the Lyman limit (LL), i.e.\
the drop in flux transmission at rest-frame wavelength $912$ \AA, corresponding to the ionization energy of hydrogen ($13.6$ eV). 
The optical depth $\mtauLL$ at this break is a defining characteristic of the 
population, with $\mtauLL = 1$ the canonical defining limit for a LLS,
corresponding to $\mNHI \approx 10^{17.2} \, \rm cm^{-2}$. 
LLS clouds trace both highly enriched gas, likely ejected into galaxy haloes by AGN or supernovae-driven winds \citep[e.g.][]{peroux06,poh+06}, 
and also gas with very low enrichment levels \citep{fop11}, 
which may represent either orphaned pockets of baryons in the IGM that have escaped enrichment by recent star formation, or the accreting gas streams that cosmological simulations predict fuel most star formation in the early Universe  \citep[e.g.][]{fpk+11,crighton16}.
The LLS distribution also influences how reionization progresses, by altering the propagation of ionizing photons through the IGM, 
affecting the morphology of ionized bubbles and the measured 21cm correlation signal
\citep{Shukla16}.
An empirical estimation of their incidence, therefore, constrains the 
processes of reionization and also informs 
models describing the accretion of cool gas onto galaxies
\citep[e.g.][]{fpk+11,fg11}.

Since the first statistical measurement of the LLS incidence \loz\
by  \cite{Tytler82}, 
with $\mloz dz$ the number of LLSs that occur at random in an interval $dz$,
there have been many measurements of the LLS distribution
\citep[e.g.][]{ssb,lzt91,key_lls}. 
The most recent use large samples of quasars over narrow 
redshift ranges, and extensively quantify systematic effects with mock observations. \citet{Prochaska_2010} measure \lox\ with a sample of 
429 quasars at $3 < z< 4.5$ from the Sloan Digital Sky Survey (SDSS). 
\citet{Ribaudo11} and \citet{OMeara13} 
assemble a combined sample of 317 quasars 
to measure \lox\ at $z < 2$, and \citet{Fumagalli13} make 
a measurement at $z=2.8$ with a survey of 61 quasars. 
\citet{Ribaudo11} showed that the evolution in the incidence rate of systems can be modelled by a power law $\mloz\ \propto (1+z)^\gamma$ with $\gamma=1.83\pm0.21$. 

Of generally greater physical interest is the quantity \lox\
with $X$ defined 
so that \lox\ is invariant if the product of the average physical size and comoving
number density of the LLS population is constant 
\citep[see section \ref{sec:formalism} of][]{Bahcall69}.
\citet{Fumagalli13} reported strong evolution in \lox\
for $\mtauLL \ge 2$ LLSs with cosmic time and introduced
a physically-motivated toy model assuming LLSs are 
produced in the haloes of galaxies in a redshift-dependent way. They find that a model where LLSs arise only in haloes with masses larger than the characteristic 
mass $M_*$ qualitatively matches the observed evolution in \lox\ at $z<4$. 
Neither of these models predict strong evolution in \lox\ at higher redshifts but, as the redshift increases towards the epoch of reionization and there are fewer sources of ionizing radiation, we expect the LLS population to increase substantially. Indeed there is some hint in the $z \sim 4$ results 
that there might be an upturn in \lox\ at $z \approx 4.5$ \citep{Fumagalli13}.

This inference, however, is tempered by the large statistical uncertainties
in the $z>4$ measurements.
Indeed, whereas the results at lower redshifts are drawn from nearly 
1,000 quasar sightlines and hundreds of LLSs, the measurements at high-$z$
have been limited by the number of existing high-quality spectra.
To date, the combined samples of \cite{Peroux03}, \cite{Songaila_2010}, 
and \cite{Prochaska_2010} form a heterogeneous set of $\approx 50$ sightlines.
In this work we present a new measurement of \lox\ for LLSs using a 
homogeneous sample of \nqso\ quasar spectra in the 
redshift range $4.4 < \mzem < 5.4$, the largest LLS survey over this redshift epoch. 
We quantify systematic errors affecting our measurement by using mock observations, and show that these effects are important, and must be accounted for to enable an accurate measure of \lox. 
Combining our results with literature measurements, we show that \lox\ at high redshift follows a power-law extrapolation fitted to the low redshift results. 
Lastly, we use these new measurements and previous work to 
constrain the \NHI\ distribution function \fnhi\ from $z \approx 2-5$.

Throughout this paper we assume a cosmology of $H_0=70\,$\kmsMpc, $\Omega_{\rm m}=0.3$ and $\Omega_\Lambda =0.7$. All distances quoted are comoving unless stated otherwise.

%
%
%
%
%
%
%
%
%

%

\section{Data}
\label{sec:data}

The quasar spectra that form the basis of our analysis are drawn
from the Giant Gemini GMOS survey of $\mzem > 4.4$ quasars 
\citep[GGG;][]{Worseck_2014}. 
The GGG survey comprises
spectra of 163 quasars selected from the Seventh Data Release of the SDSS \citep[DR7;][]{Abazajian09}. 
Only quasars with emission redshifts in the range 
$4.4 < \mzem < 5.4$ were surveyed, with a median $\mzem$ of $4.8$. This provides a sample suitable for measuring the mean free path of hydrogen-ionizing photons by averaging the Lyman continuum absorption along many quasar sightlines. For a full description of the data sample, see \cite{Worseck_2014}.

This quasar redshift range is also suitable for a statistical measure of {\it individual} Lyman limit systems: it is above the interval $2.7 < \mzem < 3.6$, where the presence of LLSs can bias the inclusion of quasars in the SDSS \citep{wp11}, 
but still at low enough redshift that IGM absorption does not prevent the identification of Lyman limit breaks.
The spectra cover $\approx 850$--$1450$\AA\ in the quasar rest-frame 
through overlapping blue and red wavelength settings of the Gemini Multi-Object Spectrographs (GMOS) on the northern and southern Gemini telescopes. 
In this work we mainly use spectra from the blue setting, which covers 
the Lyman limit and \Lya\ forest. 
The full-width-at-half-maximum (FWHM) resolution for the blue setting is $\sim 320$ \kms, 
and the typical signal-to-noise ratio (S/N) is 20 per $1.85\,$\AA\ pixel in the \Lya\ forest
at the estimated, unabsorbed continuum.

As described by \citet{Worseck_2014}, one GGG quasar spectrum has 
contaminating absorption by a nearby object on the sky 
and the sky background is therefore poorly estimated. 
It was removed from our analysis. 
Another nine of the quasars show broad absorption lines at \CIV, \NV\ and/or \OVI. 
Although we may be able to identify LLSs 
in these spectra, the broad absorption lines may introduce systematic 
effects which are absent in non-BAL spectra, so these sightlines are also removed from our analysis sample.

Table~\ref{tab:qsos} lists the set of \nqso\ quasars whose GGG spectra
were included in the LLS survey. The emission redshifts \zem\
were carefully
estimated by \cite{Worseck_2014} and are critical to defining the search
path for LLSs, as described in the following section.

\begin{table}
\begin{center}
\caption{Quasar sample for the LLS survey. All QSOs were drawn from the GGG survey \citep{Worseck_2014}. The full table is available as Supporting Information online. \label{tab:qsos}}
\begin{tabular}{lccccccc}
\hline 
Name & RA & DEC & \zem 
& $z_{\rm min}^a$ & $z_{\rm max}^b$ 
\\ 
& (deg) & (deg) 
\\ 
\hline 
J0011+1446 & 2.81349 & 14.7672 & 4.970& 4.3968 & 4.9106\\ 
J0040-0915 & 10.22772 & -9.2574 & 4.980& 4.7870 & 4.9205\\ 
J0106+0048 & 16.58017 & 0.8065 & 4.449& 3.8803 & 4.3947\\ 
J0125-1043 & 21.28927 & -10.7169 & 4.498& 4.2601 & 4.4433\\ 
J0210-0018 & 32.67985 & -0.3051 & 4.770& 4.2037 & 4.7125\\ 
J0231-0728 & 37.90688 & -7.4818 & 5.420& 5.3389 & 5.3561\\ 
J0331-0741 & 52.83194 & -7.6953 & 4.734& 4.3345 & 4.6769\\ 
J0338+0021 & 54.62211 & 0.3656 & 5.040& 4.4435 & 4.9799\\ 
J0731+4459 & 112.76304 & 44.9971 & 4.998& 4.5146 & 4.9383\\ 
J0800+3051 & 120.09590 & 30.8503 & 4.676& 4.3860 & 4.6195\\ 
J0807+1328 & 121.81300 & 13.4681 & 4.880& 4.6647 & 4.8215\\ 
\hline 
\end{tabular}
\end{center}
{$^a$}Defined for the discovery of $\tau \ge 2$ LLSs in the GGG spectra.  See the text for details.\\ 
{$^b$}Defined to be 3,000\kms\ blueward of \zem\ to avoid the quasar proximity region. \\ 
\end{table}

\section{Formalism}
\label{sec:formalism}

Determining the minimum optical depth at the Lyman limit for LLSs, \tauLL, that can be reliably determined in the GGG spectra is the first stage of defining the LLS survey. A \tauLL\ of unity
corresponds to a column density $\mNHI=10^{17.2}\,$\cmm\ and 
flux transmission $T=0.37$, while a LLS with $\mtauLL = 3$ ($\mNHI>10^{17.7}\,$\cmm, $T<0.05$) absorbs almost all the flux at the break. 
Systems with $\mtauLL \lesssim 1$ are generally referred
to as partial LLSs (or pLLSs) and are not considered in this paper \citep[see, e.g.,][]{lehner+14}.
After assessing the data quality and performing tests on mock spectra
($\S$~\ref{s:sys_mocks}), we decided to limit our analysis to LLSs
with $\mtauLL \ge 2$ in this paper.
This follows \cite{Prochaska_2010} who identified several 
biases inherent to LLS surveys and concluded that $\mtauLL = 2$
was the limit for robust analysis with low-dispersion data
of high-$z$ quasars like those in the GGG survey.  This limit has the added advantage of matching that used in many other recent measurements at $z<4$, thereby enabling direct comparison of results. An LLS with $\mtauLL \ge 2$, corresponds 
to $\mNHI \ge 10^{17.5}\,$\cmm\ and $T < 0.14$.

The survey aim, then, is to measure the incidence rate of $\mtauLL \ge 2$ LLSs, $\ell(z)_{\tau \ge 2}$. That these systems absorb at least $85\%$ of the flux at their Lyman break makes them straightforward to identify, but it also hinders the identification of lower-redshift LLSs in the same spectrum: the search path for LLSs is effectively cut short by the highest-redshift LLS removing flux at wavelengths shorter than its rest-frame $912\,$\AA. As a result, each observed LLS modifies the total search path of the survey. Previous work has shown \citep{Tytler82,Bechtold84} that 
for a LLS survey of $n_{\rm qso}$ quasars containing $m$ LLSs the maximum likelihood estimator for the incidence rate, $\ell(z)$, is then
\begin{equation}
\mloz = \frac{m}{\sum_{i=1}^{n_{\rm qso}}(z^i_{\rm max} - z^i_{\rm min})} \;\; ,
\label{eqn:loz}
\end{equation}
where $z^i_{\rm min}$ and $z^i_{\rm max}$ are the lowest and highest redshifts in the $i$th quasar spectrum where a LLS can in principle be identified. 
When a Lyman limit system cuts short the search path, 
then that LLS must be included in the resulting modified search path, 
i.e.\ $z^i_{\rm min} = z_{\rm LLS}$.
The denominator represents the integrated
redshift path of the survey where a LLS could be detected, 
often referred to as $\Delta z$. 
Equation~\ref{eqn:loz} is generally evaluated in bins of redshift and
one defines \goz, the sensitivity function, as the number of 
quasar sightlines that may be searched for LLSs at a given redshift.
It follows that $\Delta z = \int g(z) dz$ with the integral performed
over a redshift bin.

To separate the effects of cosmological expansion from any evolution in the intrinsic properties of LLSs, we convert the incidence rate as a function of redshift $\ell(z)$ to an incidence rate as a function of absorption distance $X$ (see e.g. \citealt{Bahcall69}), given by
\begin{equation}
dX = \frac{H_0}{H(z)} (1+z)^2 dz
\end{equation}
where $H$ is the Hubble parameter. The resulting incidence rate is then
\begin{equation}
\ell(X) = \frac{m}{\sum_{i=1}^{n_{\rm qso}}(X^i_{\rm max} - X^i_{\rm min})} \;\; .
\end{equation}

\section{Analysis}
\label{s:analysis}

There are two main steps required to measure \lox. 
The first is to identify  and characterize each LLS in the GGG spectra 
by strength ($\mtauLL$) and redshift (\zlls). 
The second is to define the redshift path for detecting LLSs by evaluating  \goz.
These steps are done independently although, as emphasized above, any LLSs
discovered modify the final search path used in the analysis.
In addition, for the LLS survey by \citet{Prochaska_2010}, which used SDSS spectra covering a lower redshift range than our GGG spectra, 
important systematics were found that could bias \lox\ by as much as $20$\%. We expect that similar systematics, especially the `blending bias', which we describe in more detail in the following sections, are also present 
in our GGG survey. 
In this section we first describe our method for identifying LLSs, and follow with our method for measuring the redshift path. Finally, we describe the set of simulated spectra that we use to quantify systematics, and to find the correction 
factor from our measured \lox, \rawlox, to the true \lox.

\subsection{LLS identification}
\label{s:lls_id}

LLSs were identified in each quasar spectrum by eye, using a 
custom-written tool\footnote{See the {\sc pyigm\_fitlls} script 
in the {\sc pyigm} repository,  \urlstyle{rm}\url{https://pyigm.readthedocs.io/en/latest}.}.
This tool enables a continuum template to be scaled to an appropriate level above the IGM absorption in the sightline, and for LLSs to 
be overlaid on the continuum and compared to the data. 
This by-eye approach is driven by the fact that systematics dominate
uncertainty in the analysis, particularly those due to (i) variations in continuum shape and normalization; and
(ii) stochastic line-blending with the \lya\ forest.
In future works, we intend to implement machine-learning
techniques to perform the analysis on future, large
datasets \citep[e.g.][]{garnett+17,parks+18}.

\begin{figure}
\begin{center}
\includegraphics[width=1\columnwidth]{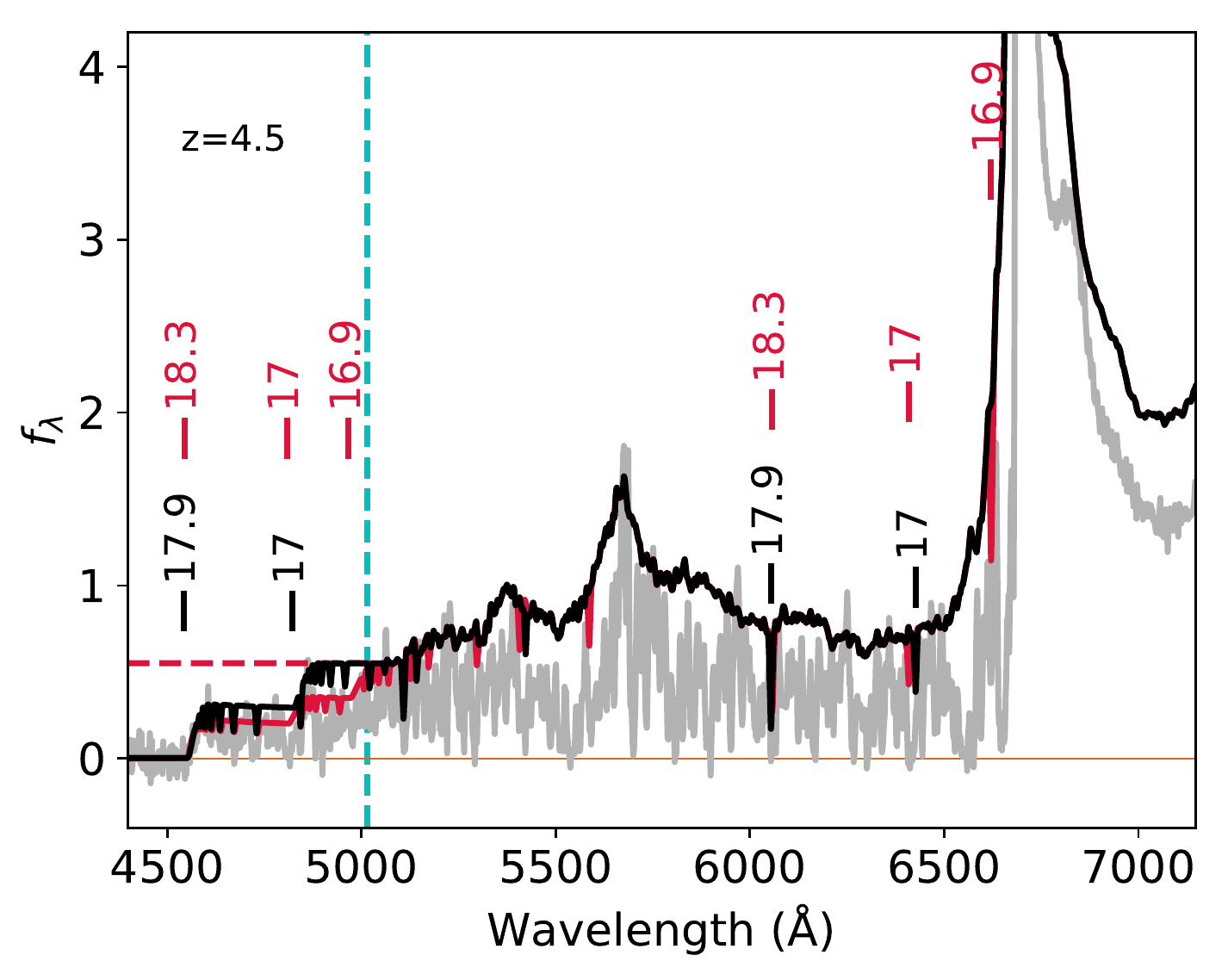}
\caption{{\label{f:spectrum} An example mock quasar spectrum (grey line) showing true and estimated Lyman limit systems. The red solid line shows the mean quasar template, which includes \Lya\ forest absorption, shifted to the redshift of this quasar and scaled such that it lies just above the flux peaks at the wavelength of the quasar Lyman limit (indicated by the vertical dashed line). The absorption from the three true Lyman limit systems in the mock line catalogue (higher, red vertical tick marks with \lNHI\ values shown) has been applied to this template. The black line shows our estimated model for Lyman limit absorption towards the quasar. Two Lyman limit systems were identified -- a system with $\mtauLL < 2$ at higher redshift, and a lower redshift $\mtauLL > 2$ system (lower, black vertical tick marks with estimated \lNHI\ values shown). Absorption from all systems identified with $\mtauLL > 0.5$ is included. In this case the highest-redshift system, a partial LLS with $\mNHI = 10^{16.9}\,$\cmm, was missed in our identification process. Our tests on such mock spectra revealed that
only LLSs with $\mtauLL > 2$ can be reliably recovered. 
}}
\end{center}
\end{figure}

For each sightline, we identified LLSs using the following steps:

\begin{enumerate}

\item  Estimate the quasar continuum level at wavelengths slightly longer than the quasar Lyman limit.  We used a template including IGM absorption, constructed by averaging all quasars in the sample (see below), normalised such that it was just above the highest flux transmission peaks at $\lambda_{\rm rest}=912$--$930$ \AA.

\item Identify the highest redshift, $\mtauLL \sim 2$ 
LL candidate, which appears as a break in the continuum shortwards of the quasar Lyman limit.  Insert a model for the optical depth for this system into the template continuum. Check the \Lya\ and other Lyman series lines for the system, adjusting its redshift to improve agreement with these lines.

\item Alter \NHI\ until the model matches the observed drop in flux at the absorber's Lyman limit.

\item Adjust the Doppler broadening parameter ($b$) of the absorber to best match the equivalent width of the Lyman series lines.

\item Repeat steps $2$--$4$ for all discernable
LLSs in the spectrum, moving from the highest to the lowest redshift candidate.

\end{enumerate}
Figure \ref{f:spectrum} illustrates this process using a mock spectrum (described in Section \ref{s:sys_mocks}).

The continuum template was generated by shifting all of the non-BAL GGG spectra to rest wavelengths, normalising in the rest wavelength range $1250$--$1280$ \AA\ (masking the sky absorption feature at $7590$--$7660$ \AA) and averaging them. Below the rest-frame quasar Lyman limit, the presence of Lyman limit systems suppresses the template below the mean IGM flux level, and so we assume a flat continuum in $F_\lambda$ at wavelengths shorter than $\lambda_{\rm rest} = 930$ \AA. We did not use metal lines to refine redshifts, because in most cases they were not strong enough to be detected at the 
S/N and resolution of our spectra. 

%

Inevitably there are some subjective criteria that determine whether a drop in flux 
in a quasar spectrum is flagged as a LLS. Therefore two of the authors (NC and JO) each identified LLS in every GGG spectrum, enabling us to assess how the identification criteria described above affect the recovered LLS statistics. As an initial validation of our search procedure, we searched for LLSs in a smaller sample of eight spectra before starting the full search. These eight spectra were selected from the real GGG spectra to show no strong Lyman limit absorption within $\delta z=0.3$ of the quasar Lyman limit, and thus create a set of diverse strong absoprtion-free continua. Between zero and two synthetic LLSs were then inserted into each spectrum, with their \NHI\ and redshift values drawn at random in the ranges 
$\mNHI=10^{\text{16.5--17.8}}$ \cmm, 
$z_{\rm QSO} - 0.3 < \mzlls < z_{\rm QSO}$, 
and with $b$ set to $20$\,\kms. Each author searched these spectra for LLSs, and we compared the recovered LLS parameters (\NHI, \zlls)
with the known parameters. This test showed that in most cases the 
$\mtauLL \ge 2$
LLS were recovered correctly, with \NHI\ and redshift close to the known values.  
We therefore concluded that our search method was sound, and continued fitting all the spectra using these criteria. In Section \ref{s:sys_mocks} we describe more quantitatively how well our search method recovers $\mtauLL > 2$ systems.
Table~\ref{tab:all_lls} lists all of the LLSs identified in the 
full sample by NC.  The final set of LLSs used to estimate
\loz\ is drawn from this list.
\begin{table}
\centering
\caption{All the LLSs recovered from the GGG spectra by author NC. 
The full table is available as Supporting Information online. \label{tab:all_lls}}
\begin{tabular}{lccccc}
\hline 
RA & DEC & \zlls & \lNHI & $b$ 
\\ 
(deg) & (deg) & & & (km/s) 
\\ 
\hline 
10.22772 & -9.2574& 4.78713 & 17.90 & 40.0\\ 
10.22772 & -9.2574& 4.97083 & 17.20 & 20.0\\ 
16.58017 & 0.8065& 4.23517 & 17.35 & 50.0\\ 
16.58017 & 0.8065& 4.06464 & 17.45 & 20.0\\ 
21.28927 & -10.7169& 4.18351 & 20.20 & 70.0\\ 
21.28927 & -10.7169& 4.26017 & 17.30 & 20.0\\ 
21.28927 & -10.7169& 4.23090 & 17.30 & 20.0\\ 
32.67985 & -0.3051& 4.55488 & 17.35 & 40.0\\ 
32.67985 & -0.3051& 4.27720 & 17.15 & 20.0\\ 
37.90688 & -7.4818& 5.33901 & 19.80 & 60.0\\ 
37.90688 & -7.4818& 4.88711 & 20.70 & 20.0\\ 
\hline 
\end{tabular} 
\end{table}

\subsection{Defining the redshift path}
\label{s:zpath}
We define the redshift path for detecting LLSs in one of our spectra
in the following way. The lower limit for the path is determined from the bluest wavelength, $\lambda_{\rm min}$\,[\AA], where the smoothed S/N -- defined as the flux-to-error ratio smoothed by a 20-pixel boxcar filter -- is $\ge$2 per pixel. 
We consider this a reasonable compromise between maximising the available redshift path length and enabling a confident detection of $\mtauLL \ge 2$ LLSs.
The lower limit to the search path for that quasar is then $\mzmin = \lambda_{\rm min}/912{\rm \AA} - 1$.
If the Lyman limit of a LLS caused the S/N to drop below this
threshold, then this definition led to the system being excluded from
the redshift path, greatly suppressing the number of LLS in the
survey. To prevent this artificial effect, we check whether there is
an observed LLS with redshift $z_{\rm min} - 0.3 < z < z_{\rm min}$, and
where such a system is present we set $z_{\rm min}$ at the LLS
redshift, and include that system in the search path if it has
$\tau_{rm LL} > 2$.
The upper search limit for the search path, \zmax\
is taken to be $3000$\,\kms\ bluewards of the quasar redshift. 
This offset is chosen to avoid the proximity zone of the quasar which
may be biased by gas clustered around the host galaxy 
\citep{QPQ2} 
and/or the quasar radiation field \citep{claude07}.

\begin{figure}
\begin{center}
\includegraphics[width=0.95\columnwidth]{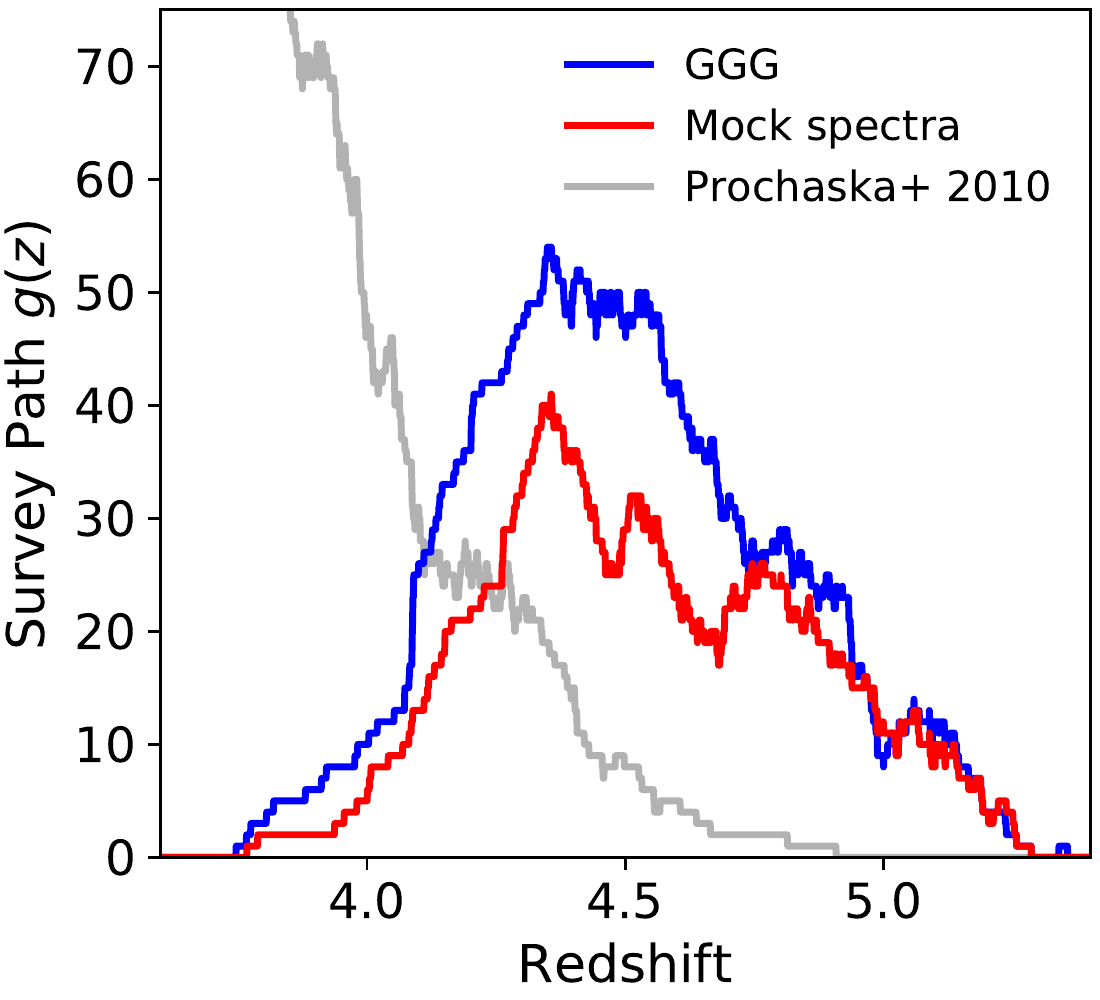}
\caption{{\label{f:zpath} The redshift path for detecting LLS, $g(z)$,  
in our GGG survey (blue), and in the mock sightlines (red). The total paths for the real and mock data differ because of their different distribution of Lyman limit systems. The light grey curve shows $g(z)$ from a previous high-redshift LLS survey with SDSS spectra \citep{Prochaska_2010}. 
The GGG and mock $g(z)$ assume a path truncated at $3000\,$\kms\ from the quasar redshift, and a minimum smoothed S/N threshold of 2 per pixel. See Section \ref{s:zpath} for further details.%
}}
\end{center}
\end{figure}

We checked the effect of changing these parameters over reasonable ranges, and found that they do not change the \lox\ measurement by more than the final $1\sigma$ error.  
Table~\ref{tab:qsos} lists the \zmin\ and \zmax\ values for each quasar,
and Fig.~\ref{f:zpath} shows the resulting sensitivity function, \goz.
The decline in \goz\ at $z>4.5$ follows the decline in the number of 
quasars in the sample at such redshifts.  The decline 
at $z<4.3$, however, is a complex convolution of the GGG quasar
redshift distribution, the GGG observing strategy,
and the incidence of LLSs which alter \zmin.

\subsection{Assessment of systematic uncertainties using mock spectra}
\label{s:sys_mocks}

Our initial check that we could successfully recover LLSs (see Section \ref{s:lls_id}) also indicated the presence of systematic errors in the recovered LLS parameters, caused by the low spectral resolution and strong IGM absorption at high redshift. This is not surprising -- \citet{Prochaska_2010}, for example, identified several possible systematic effects introduced when identifying LLSs in SDSS quasar spectra. Their survey used higher resolution, generally noisier spectra, and was at a lower redshift, but we expect many of the systematic effects to also be present in our GGG survey. To assess these systematic errors we generated a sample of mock spectra, one corresponding to each real GGG 
spectrum, and applied simulated IGM absorption to them. These spectra were then searched for LLSs in the same way as the real spectra by the same two authors (NC and JO). We compared the parameters of LLSs recovered by our search to the known parameters of the LLSs inserted into the mocks to quantify any systematic uncertainties in our LLS identifications. Appendix \ref{a:mocks} discusses our IGM simulations and how we generated the mocks.

\begin{figure}
\begin{center}
\includegraphics[width=1\columnwidth]{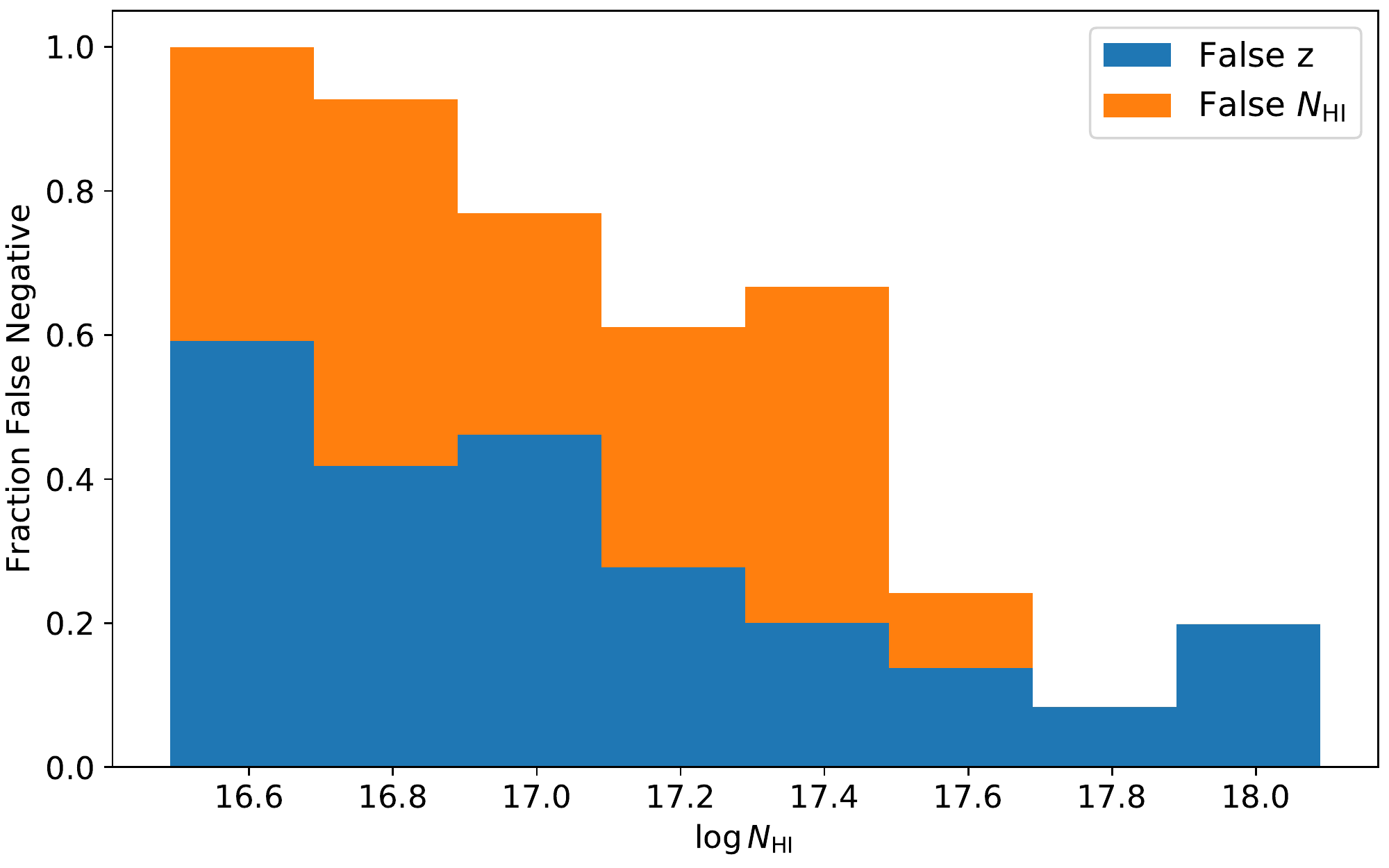}
\caption{{\label{f:false} 
Incidence of false positives, true LLSs in mock spectra that were
missed or mis-identified by the human.  Orange indicates the fraction of
cases where the human mis-estimated the \NHI\ value by more than 0.3\,dex.
The blue histogram are cases with $\delta v \ge 3000$\kms.
For $\mNHI > 10^{17.5}$\cmm, corresponding to $\mtauLL \ge 2$,
the total incidence of false negatives falls below an acceptable
$20\%$.
}}
\end{center}
\end{figure}

By searching these mock spectra we found that LLSs with $\mtauLL \ge 2$ could be recovered with high confidence. 
Fig.~\ref{f:false} shows the incidence fraction of false negatives from the mock analysis with 
a false negative defined as a true LLS that was mis-identified by the human.
We show separately: 
(i) false negatives in redshift, defined as a true LLS where
the nearest human-identified LLS was offset by more than 3,000~\kms, and
(ii) false negatives in \NHI, where the estimate \NHI\ value was offset by
at least 0.3\,dex (where all \NHI\ values were capped at a maximum of 17.8\,dex).
These are shown as a function of true \NHI.
The combined false negative rate is 100\%\ for the lowest
\NHI\ LLSs injected, i.e.\ none were correctly recovered.
Weaker systems were more difficult to identify, as the relatively shallow depression of flux they produce below the quasar Lyman limit is easy to overlook completely or misinterpret as IGM absorption or quasar continuum fluctuations. 
The results show further that at $\mNHI \approx 10^{17.5}$\cmm, the
rate of false negatives drops markedly.

\begin{figure}
\begin{center}
\includegraphics[width=1\columnwidth]{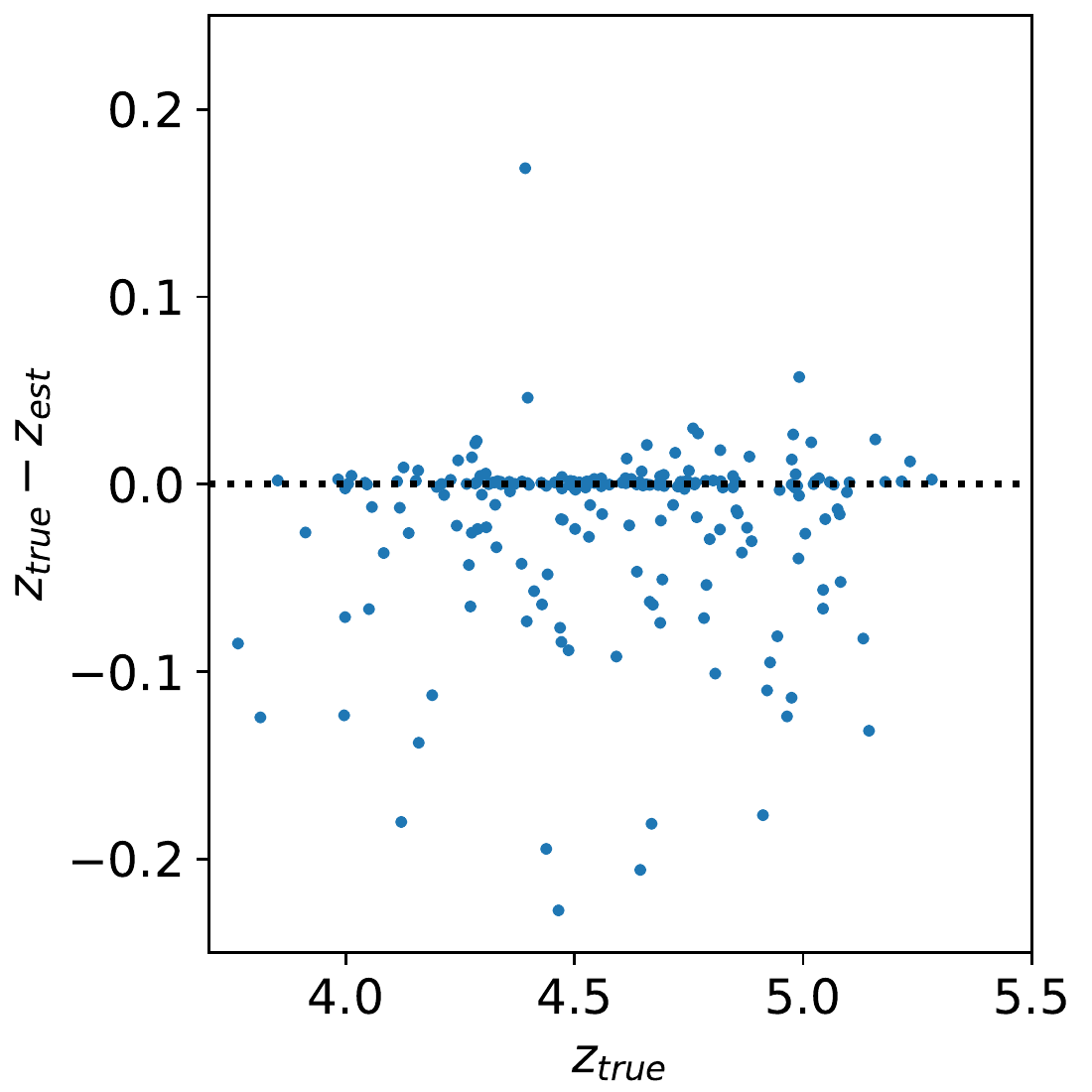}
\caption{{\label{f:dz} Difference between the true and estimated redshift for Lyman limit systems in the mock spectra for LLSs identified by one of the authors (NC). Results are shown for systems with true $\mtauLL>2$. Large differences ($\delta z > 0.1$) are usually due to an incorrect match, where the true LLS has been missed and instead matched to a LLS at a different redshift. 
}}
\end{center}
\end{figure}

LLSs were most straightforward to identify in quasars with a single, $\mtauLL > 2$ system. More complex sightlines with many, closely-spaced partial LLSs were more difficult to recover accurately. For these complex sightlines, we sometimes identified a single strong LLS to explain the absorption of several lower $\mNHI$ LLSs, and the redshift difference between the input and recovered LLS could be quite large ($\delta z\sim 0.1$). To quantify how well we are able to recover the LLS redshift from the mock spectrum, we take each `true' LLS with $\mtauLL > 2$ in the mocks (described in more detail below) and match it to the nearest guessed LLS in redshift in that spectrum. The results are shown in Fig.\ \ref{f:dz}. For most LLSs, the redshift uncertainty is very small, $\left|\delta z\right| < 0.02$, but the distribution has broad tails out to $\left|\delta z\right| \pm 0.2$. This suggests we can reliably identify strong LLSs, and so make a robust \lox\ measurement. 

\subsection{Finding the true \lox}
\label{s:true_corr}

We take the `true' \lox\ of the mock skewers to be the line density over the redshift range $3 < z < z_{\rm qso}$ averaged over all quasars. Due to the way we assign LLSs to the simulated skewers (see Appendix \ref{a:mocks}), they often have very small separations, $\delta v < 500$\,\kms. In high resolution spectra, LLS components are only judged as belonging to distinct systems if they are separated by more than $\sim 500$\,\kms\
\citep{prochaska+15},
which is the circular velocity for high mass ($10^{12.8} M_\odot$) dark matter haloes at $z=4.5$, and corresponds to the largest range typically observed for metal-lines associated with strong LLSs and DLAs \citep[e.g.][]{Peroux03}. Therefore, we merge systems with separations less than $\left|\delta v\right| < 500$\,\kms, assigning a new redshift given by the mean of the contributing components' redshifts, weighted by their column densities. This gives the true \lox\ we compare to the measured values from the mock spectra. The resulting true \lox\ is not sensitive to the precise merging scale; 
if we use $|\delta v| < 750$ instead of $500$\,\kms, the 
true \lox\ changes by less than 2\%.

\begin{figure}
\begin{center}
\includegraphics[width=0.92\columnwidth]{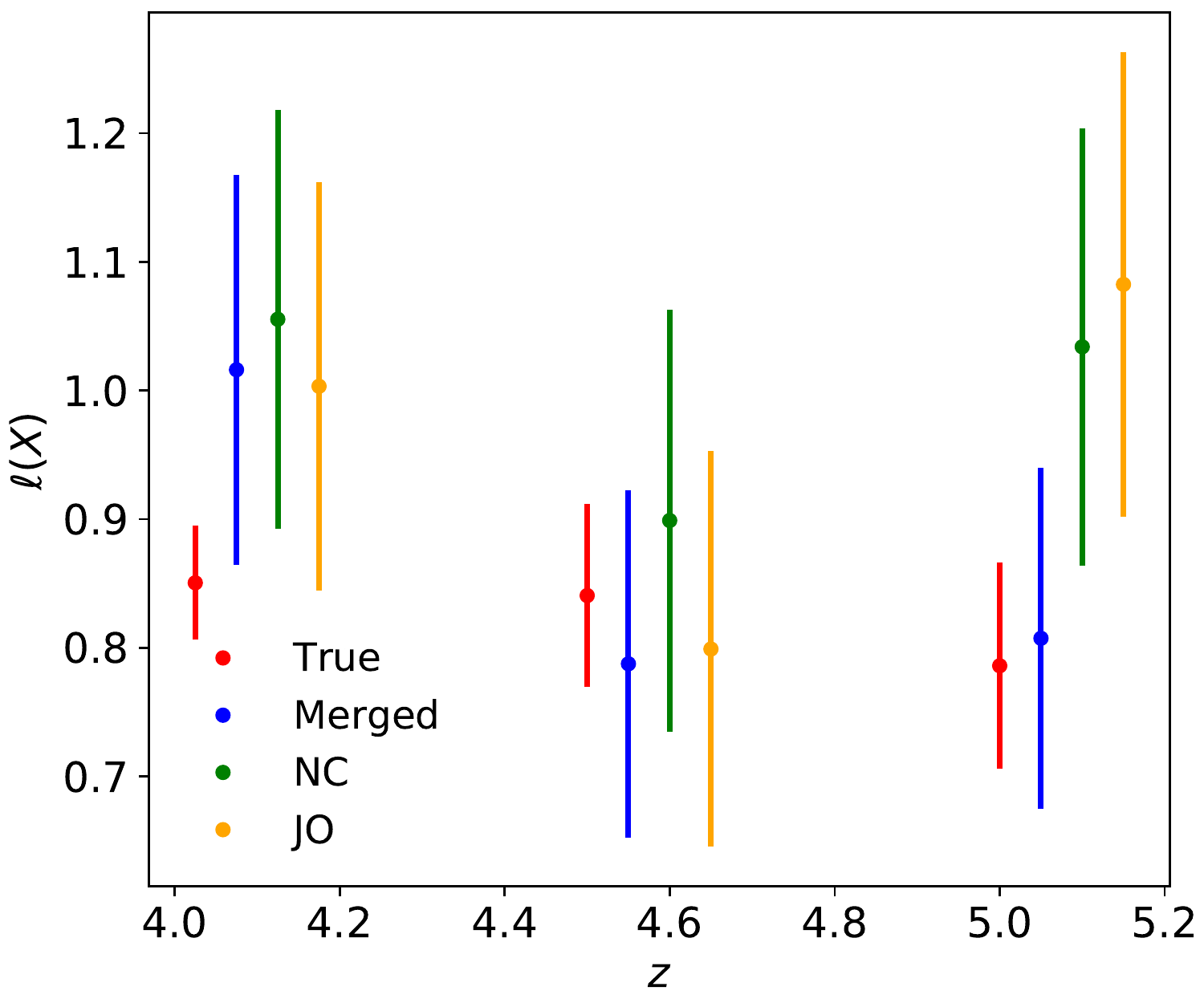}
\caption{{\label{f:corr} Comparison of the observed and true \lox\ from the analysis of mock spectra in three redshift bins. Red points show the true value, after
summing all components with separations less than 500\,\kms\ into a single LLS (see text). Error bars are $1\sigma$ Poisson uncertainties given the number of systems in each bin. 
\lox\ inferred from the LLSs identified by NC and JO are shown by green 
and orange points. 
These values are consistent within the 1$\sigma$ uncertainties plotted, suggesting that any subjective criteria used to identify LLSs that differ between authors do not significantly impact our analysis. The blue points show the true \lox\ value, where we introduce a further merging of systems to approximate the tendency to merge nearby weaker systems into a single stronger system when recovering LLSs in the mocks (see text). The  lowest redshift bin and the uncertainties in all bins more closely matches the inferred \lox\ from both authors, which suggest this merging is an important systematic effect.%
}}
\end{center}
\end{figure}

Figure \ref{f:corr} compares \lox\ derived from the human analysis   
of the mock spectra with the true evaluation discussed above for $\mtauLL \ge 2$.
These results suggest that \lox\ is overestimated by $\sim20$\% in the lowest of three redshift bins, and by $\sim 40\%$ in the highest redshift bin. By inspecting our fits to the mock spectra and overplotting the known LLS positions (see Fig.\ \ref{f:spectrum} for an example), we determined that the main systematic causing this excess is the `blending bias', as described by \citet{Prochaska_2010}. This refers to the effect from nearby, partial LLSs (with $\mtauLL < 2$) that tend to be merged together when identified by their Lyman break alone, and thus are incorrectly identified as a single $\mtauLL > 2$ LLS. This merging occurs due to confusion in identifying the redshift of a LLS: often, all the Lyman series lines have a relatively low equivalent width, and so the only constraint on the LLS redshift is the LL break. The position of the break can be affected by overlapping IGM absorption, and by the velocity structure of the LLS, which is difficult to measure accurately using low-resolution spectra like the GMOS data employed here.

\begin{figure}
\begin{center}
\includegraphics[width=0.75\columnwidth]{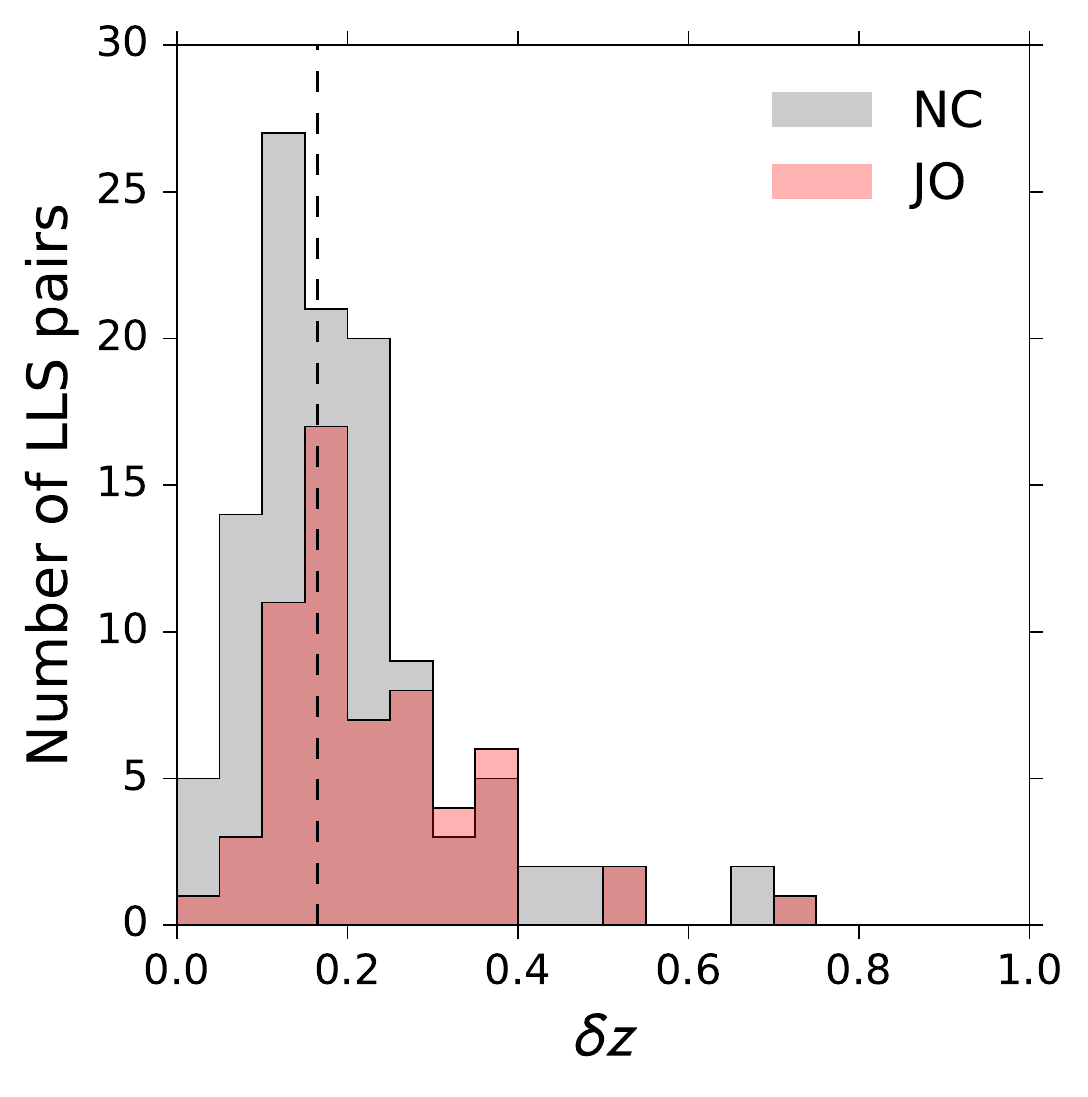}
\caption{{\label{f:dz_LLS_one_sightline} Distribution of absolute redshift 
differences, $\delta z$, between pairs of LLSs in the same sightline for the GGG spectra. 
These include systems with $\mtauLL < 2$, in addition to $\mtauLL > 2$ systems. Results from two authors, NC and JO, are shown. When measuring \lox\ directly from the mock line catalogue we adopt a merging scale for neighbouring LLSs of $9000\,$\kms\ ($\delta z \sim 0.165$ at $z=4.5$, shown as the vertical line in the plot), corresponding to the peaks of the histograms.%
}}
\end{center}
\end{figure}

To further illustrate the importance of this systematic effect, 
we apply a simple merging scheme to the true list of LLSs, and compare \lox\ for this merged list to our recovered \lox.  To do this merging, we step through each LLS with $\mtauLL > \tau_1$ in the true list (described above), from lowest to highest redshift, and test whether there is another $\mtauLL > \tau_1$ system 
within 9000\,\kms (see below and Fig.~\ref{f:dz_LLS_one_sightline}
for the justification of this value). 
If one is found, we merge the two systems by adding their column densities, using the redshift of the highest redshift system for the resulting single combined system. 
This process is repeated, with the new merged system replacing the previous 
two separate systems, and continued until every system 
with $\mtauLL> \tau_1$ has been tested in a sightline. $\tau_1=1$ corresponds to the optical depth of a LLS which produces an appreciable Lyman break, detectable in the GMOS spectra. Using our mocks, we also explore the effect of setting $\tau_1 = 0.5$ or 1.5 instead and find that, while it has a negligible effect for the highest and lowest redshift bins in Fig.\ \ref{f:corr}, it can introduce an uncertainty of up to 8\% in \lox\ for the central redshift bin. This is significantly smaller than the final statistical uncertainty on \lox\ for this bin, but it is large enough that we include its contribution in the final uncertainty budget. 

To estimate the merging scale, we measured the distribution of redshift differences between LLSs that we observe towards sightlines showing more than one LLS. Fig.\ \ref{f:dz_LLS_one_sightline} shows these distributions for both authors who identified LLSs. 
Without the blending bias, one expects a roughly flat distribution at 
small $\delta z$ that then drops with increasing $\delta z$ because the
presence of one strong LLS precludes the detection of any other at lower $z$.
The low number of pairs with $\delta z < 0.1$, therefore, must be related to the
blending bias.
We adopt the peak in these distributions as the smallest scale at which we can reliably separate nearby LLSs. The peak for both authors is at $\delta z \sim 0.17$, equivalent to 9270\,\kms\   at $z=4.5$, and we therefore adopt a merging scale of 9000\,\kms. Using a merging scale of 6000 or 12000\,\kms\ instead of 9000\,\kms\ changes the inferred \lox\ by less than 3\%. Finally, we define a redshift path for this merged list in a similar way for the mock spectra, by truncating the redshift path in a sightline when it encounters a strong $\mtauLL>2$ LLS.

With this new, merged list of LLSs in the mock spectra, we then make a new measurement of \lox, shown by the blue points in Fig.\ \ref{f:corr}. While these blue points do not precisely match the directly measured \lox\ values, they do show similar systematic offsets -- the first and last bins have an increase in \lox\ and there is little change in the central bin. We consider this merging effect, which truncates the search path and merges weaker LLSs into $\mtauLL > 2$ systems, to be the main systematic error to correct in our analysis. 
In the following section we apply the correction derived 
from this mock spectrum analysis to the real GGG sightlines, 
and report the final, corrected \lox\ values.

%

\subsection{\lox\ measurements}

\begin{figure}
\begin{center}
\includegraphics[width=1\columnwidth]{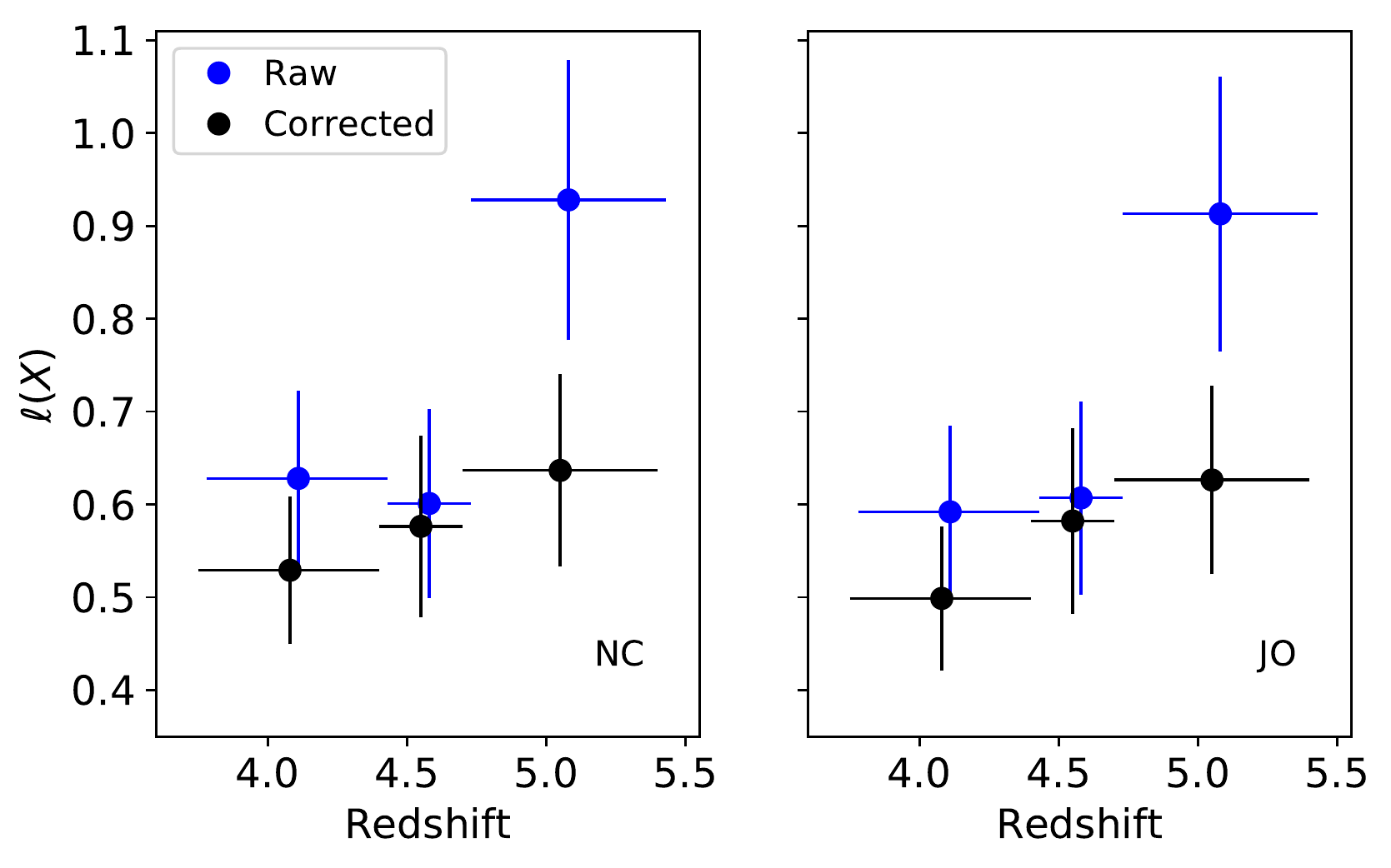}
\caption{{\label{f:lX_corr} \lox\ before and after applying the correction factors inferred from mock spectra. The raw, uncorrected \rawlox\ values are shown 
in blue, offset slightly in redshift for clarity. The left panel shows the results from one author (NC), and the right panel for another (JO). The corrected (true) \lox\ values are 
shown in black. Error bars represent the $1\sigma$ Poisson uncertainty. The corrected results for JO and NC are in good agreement and the largest discrepancy between the corrected values for each author is in the lowest redshift bin (a 4\% difference). This discrepancy is much smaller than the final statistical error on \lox.
}}
\end{center}
\end{figure}

Table \ref{tab:analy_lls} lists the final set of $\mtauLL \ge 2$ LLSs used for the statistical analysis of \lox. We have measured \lox\ in three redshift bins, 
$z=3.75$--4.4, 4.4--4.7, 4.7--5.4, chosen to
have roughly equal numbers of LLSs.
The raw \rawlox\ measurements from analysis by NC and JO are shown
in Figure~\ref{f:lX_corr}.  
The two sets of \rawlox\ measurements are in excellent agreement;
the largest offset is only $\approx$4\%.  We have then corrected
these \rawlox\ values based on the mock spectrum analysis
described in Section \ref{s:true_corr}.
Table \ref{tab:lzlX} summarizes the \lox\ measurements
including the corresponding values for \loz.
We have also estimated $\mloz$ across the entire
survey ($z \approx 4.4$) to be $\mloz = \lzfull \pm \slzfull$. 

\begin{table}
\centering
\caption{LLSs recovered from the GGG sample that are used for the statistical analysis. The full table is available as Supporting Information online. \label{tab:analy_lls}}
\begin{tabular}{cccc}
\hline 
RA & DEC & \zlls & $\log \mNHI$ 
\\ 
(deg) & (deg) 
\\ 
\hline 
10.22772 & -9.2574& 4.78713 & 17.90\\ 
37.90688 & -7.4818& 5.33901 & 19.80\\ 
52.83194 & -7.6953& 4.33463 & 17.80\\ 
112.76304 & 44.9971& 4.51475 & 17.90\\ 
120.09590 & 30.8503& 4.38613 & 19.00\\ 
121.81300 & 13.4681& 4.66482 & 17.80\\ 
125.55146 & 16.0769& 3.91159 & 19.70\\ 
126.22510 & 13.0381& 5.02951 & 17.90\\ 
129.23253 & 6.6846& 4.20632 & 17.90\\ 
129.83555 & 35.4165& 4.42728 & 18.00\\ 
131.63137 & 24.1857& 4.50460 & 17.90\\ 
\hline 
\end{tabular} 
\end{table}

\begin{table*}
\centering
\caption{Raw and corrected LLS incidence rates in the GGG spectra based on one author's LLS identifications (NC). \label{tab:lzlX}}
\begin{tabular}{ccccccccccccc}
\hline 
Redshift bin & $m$ & $\Delta z$ & $\Delta X$ & \rawloz & \loz & $-1\sigma$ & $+1\sigma$ & \rawlox & \lox  & $-1\sigma$ & $+1\sigma$
\\ 
\hline 
3.75--4.40 & 42 & 15.29 & 63.22 &2.75 & 2.21& 0.34 & 0.40& 0.66 & 0.54& 0.08 & 0.10\\ 
4.40--4.70 & 31 & 13.15 & 56.12 &2.36 & 2.20& 0.39 & 0.47& 0.55 & 0.52& 0.09 & 0.11\\ 
4.70--5.40 & 36 & 9.28 & 40.89 &3.88 & 2.95& 0.49 & 0.58& 0.88 & 0.67& 0.11 & 0.13\\ 
\hline 
\end{tabular} 
\end{table*}

\begin{figure}
\begin{center}
\includegraphics[width=1\columnwidth]{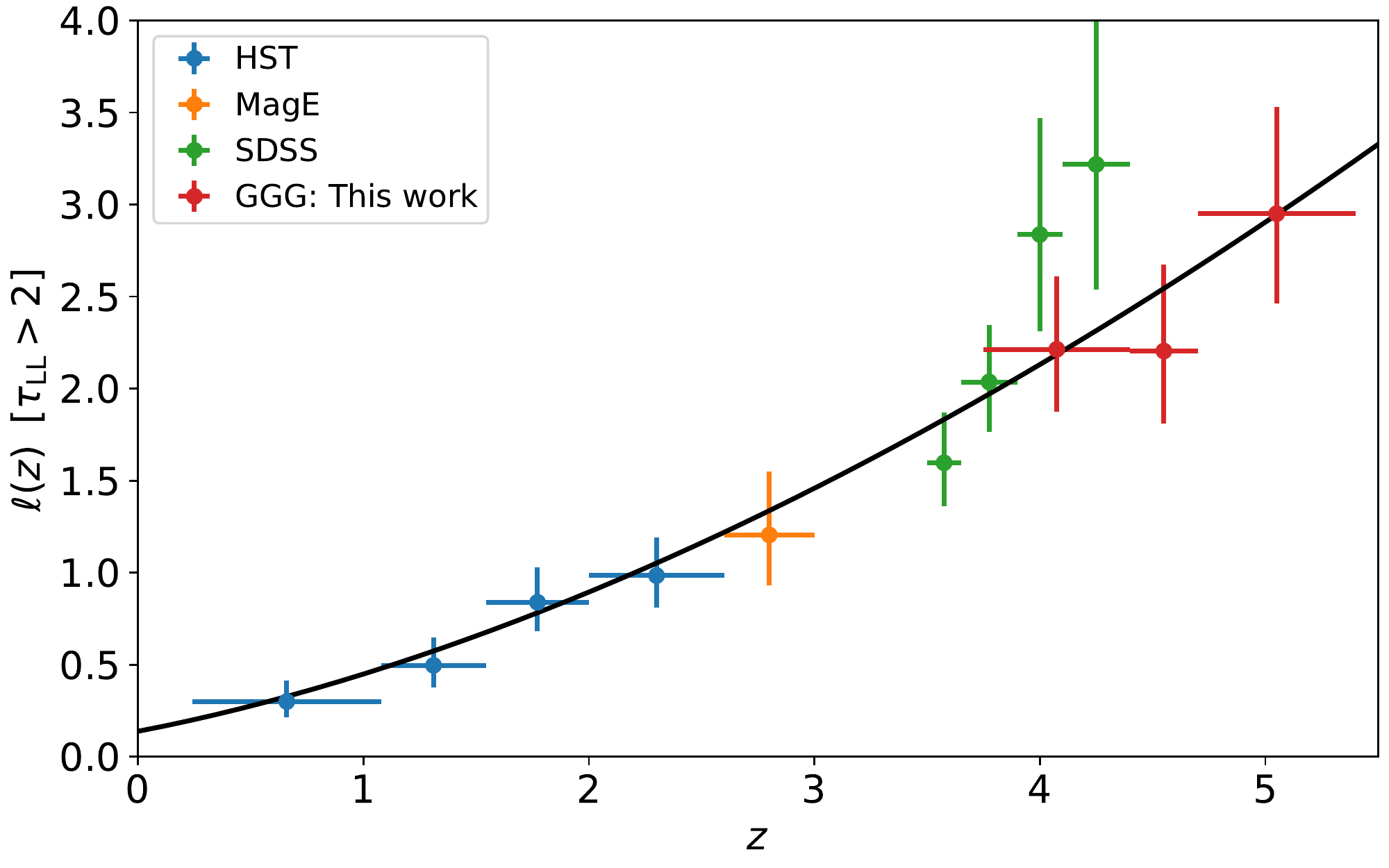}
\caption{{\label{f:loz} 
Binned evaluations of \loz\ for LLSs with $\mtauLL \ge 2$
from this study (red points) and a series of measurements
from the literature re-analyzed using the {\sc PYIGM}
package:  \citep[{\it HST};][]{Ribaudo11,OMeara13}
\citep[MagE;][]{Fumagalli13}, \citep[SDSS;][]{Prochaska_2010}.
Uncertainties are Wilson score estimates.
Overplotted on these data is the best-fit power-law 
$\mloz = \ell_* [(1+z)/(1+z_*)]^\alpha$
with $z_*=3$, 
$\ell_* = \lstar$, and $\alpha = \aval$.
This model appears to be an excellent description of the data
over the $\approx 10$\,Gyr considered.
}}
\end{center}
\end{figure}

Previous work \citep[e.g.][]{Ribaudo11} 
has shown that \loz\ measurements are well described 
by a power-law of the form 
$\mloz = \ell_* [(1+z)/(1+z_*)]^\alpha$
where $z_*$ is an arbitrary pivot redshift.
We compare this model against the combination of our GGG results and the {\it HST} datasets of 
\cite{Ribaudo11} and \cite{OMeara13}, the MagE survey
by \cite{Fumagalli13}, and the SDSS results from
\cite{Prochaska_2010}\footnote{All of the
LLS lists, sensitivity functions, and code to calculate \loz\ 
are available in the {\sc pyigm} repository at \urlstyle{rm}\url{https://github.com/pyigm/pyigm}\,.}.  
We adopt the maximum likelihood approach to fitting 
the LLS surveys, described in \cite{Prochaska_2010},
and recover $\ell_* = \lstar$ and $\alpha = \aval$
(68\% confidence intervals) with $z_* = 3$.
Binned measurements of \loz, assuming binomial statistical
uncertainties, are compared against this best-fit model in
Fig.\ \ref{f:loz}.  We have performed a Kolmogorov--Smirnov test 
comparing the cumulative
distribution function of $z_{\rm LLS}$ from the observations
and that derived from the convolution of the \loz\ best-fit and 
the survey sensitivity functions.  We recover an acceptable 
probability of only 23\%\ that the null hypothesis is ruled
out.

\begin{figure}
\begin{center}
\includegraphics[width=1\columnwidth]{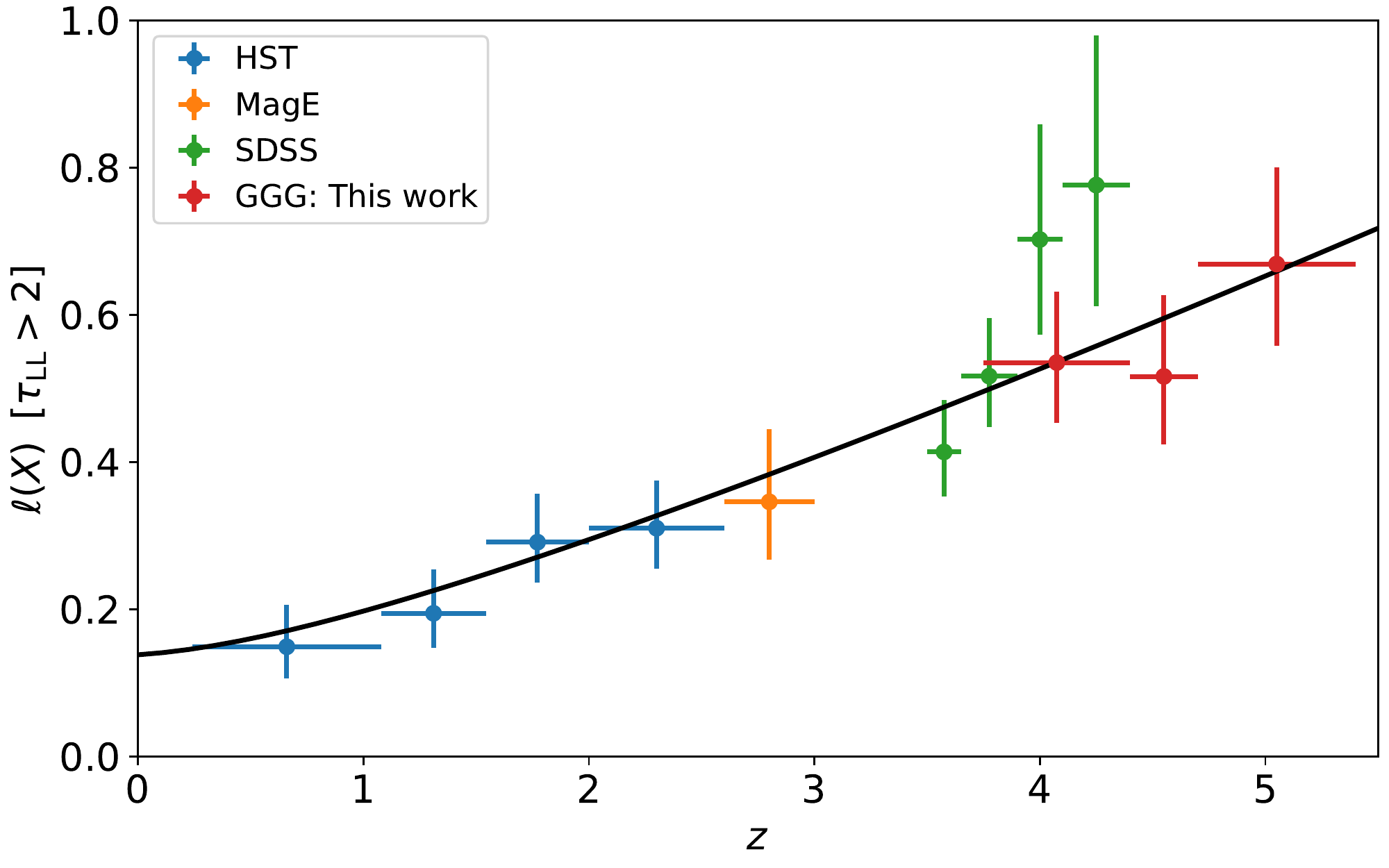}
\caption{{\label{f:lX} Same as for Fig.~\ref{f:loz}
but for \lox.  The curve shown is the fit to \loz\ but translated
to \lox\ with our adopted cosmology.
}}
\end{center}
\end{figure}

Fig.\ \ref{f:lX} presents the final \lox\ measurements for the GGG 
sample using the results from one author (NC) from Fig.\ \ref{f:lX_corr}. 
These are shown together with evaluations of \lox\ for
$\mtauLL > 2$ systems from previous, lower redshift measurements. 
There is no evidence for a sudden change in the LLS incidence rate at $z\sim4$--5, which has been speculated about previously 
\citep[e.g.][]{Fumagalli13,Crighton15}
based on the highest redshift points from 
the SDSS analysis of \citet{Prochaska_2010}.
Indeed, the two highest redshift bins from the SDSS
sample have large uncertainties
and are consistent with our results within the combined uncertainties.
Furthermore, we suspect a correction for the blending bias
should be applied to those SDSS measurements, in a similar fashion performed for the GGG results here.


\subsection{Clustering of LLSs}
\label{sec:cluster}

Clustering of LLSs can potentially introduce a systematic effect where the 
measured \rawlox\ significantly underestimates the true \lox. This is because of the survival nature of the measured 
\rawlox, where a single LLS in a sightline prevents detecting any lower redshift LLS. Therefore, only the highest redshift component of a cluster of several LLSs is measured.  
In principle this can suppress \lox\ by a factor comparable to the number of LLSs per cluster, an effect that has not (to our knowledge) 
been quantified in previous \lox\ measurements \citep[as noted by][hereafter \citetalias{pmo+14}]{pmo+14} 
Indeed, \citet{qpq6} reports a large clustering length-scale $r_0 > 10 \, h^{-1}$\,Mpc
for the LLS--quasar cross-correlation function, suggesting a high degree
of clustering near massive haloes.  An ongoing survey with 
pairs of quasar spectra will provide the auto-correlation function
for LLSs in a random distribution of environments (Fumagalli et al., in preparation).

\begin{figure}
\begin{center}
\includegraphics[width=0.84\columnwidth]{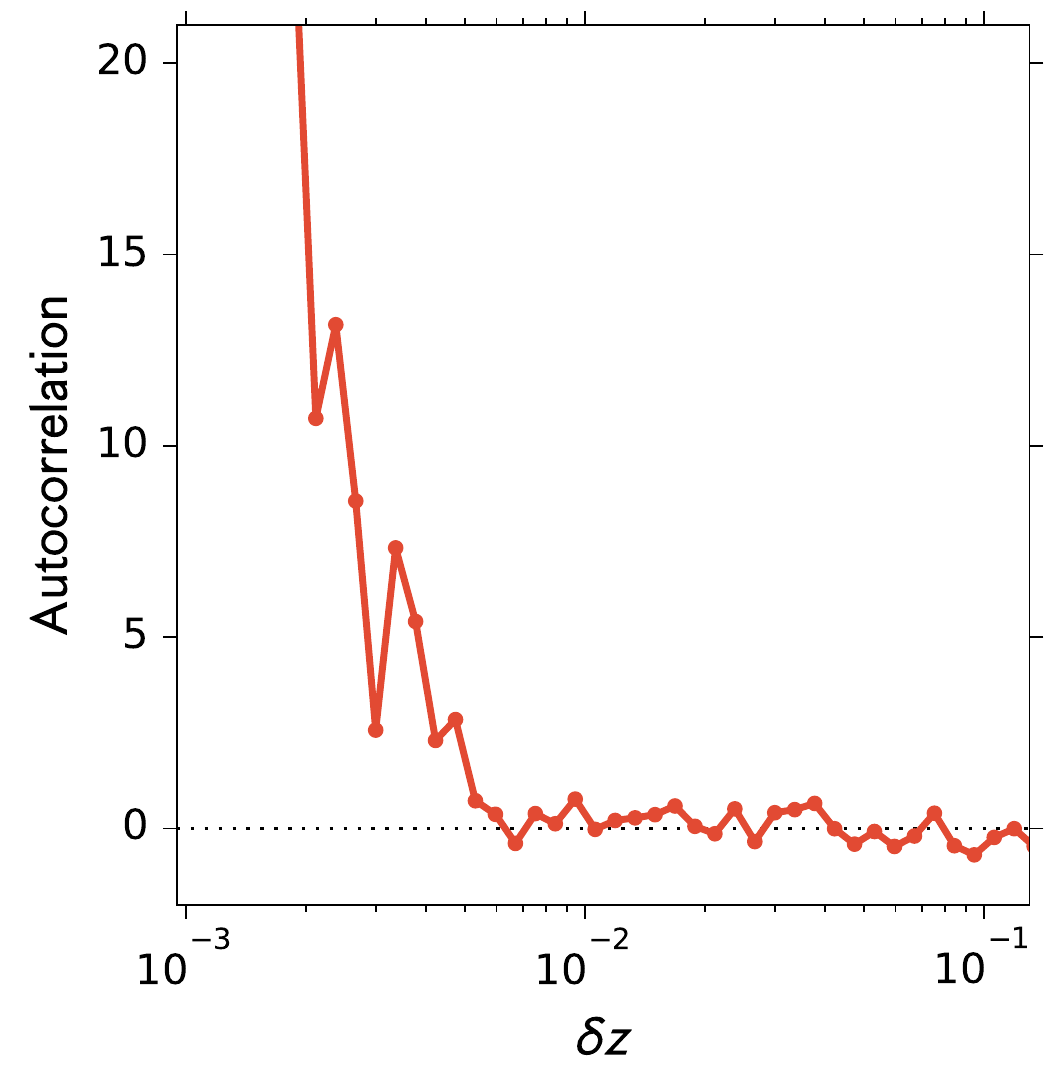}
\caption{{\label{f:acorr} The autocorrelation of LLSs in the simulated sightline skewers (see Appendix \ref{a:mocks}) as a function of redshift separation. There is no strong correlation beyond $\delta z = 0.01$, and so we do not expect LLS clustering to affect our \lox\ measurement.%
}}
\end{center}
\end{figure}

We use our simulated `skewers' (see Appendix \ref{a:mocks}) to estimate the importance of this clustering in the simulated spectra. We measure the autocorrelation of LLSs with $\tau_\mathrm{LL} > 2$ in the simulated skewers as a function of redshift separation. The results are shown in Fig.\ \ref{f:acorr}. Correlation is only significant at scales $\delta z<0.01$ (550\,\kms\ at $z=4.5$). Since these scales are much less than the merging length described above (9000\,\kms), they imply that clustering will not not have a significant effect on our measured \lox.
%
%
However, we emphasise that this conclusion should be tested with simulations which more accurately reproduce the distribution of LLSs and include radiative transfer effects.

\begin{table*}
\centering
\caption{Constraints on \fnhi\ at $z \sim 4.5$ \label{tab:z5_fn_constraints}}
\begin{tabular}{cccccccccc}
\hline 
Constraint & $z$ & $\log \mNHI^{\rm min}$ & $\log \mNHI^{\rm max}$ & Value & $-1 \sigma$ & $+1\sigma$ & Reference\\ 
\hline 
\lmfp & 4.56 &11.0 & $\infty$&22.20&2.3 & 2.3& Worseck+14\\ 
\lmfp & 4.86 &11.0 & $\infty$&15.10&1.8 & 1.8& Worseck+14\\ 
\lmfp & 5.16 &11.0 & $\infty$&10.30&1.6 & 1.6& Worseck+14\\ 
log \fnhi & 4.40 &20.300 & 20.425& -22.12& 0.24 & 0.21& Crighton+15\\ 
&&20.425 & 20.675& -22.13& 0.18 & 0.15& Crighton+15\\ 
&&20.675 & 21.075& -22.51& 0.18 & 0.13& Crighton+15\\ 
&&21.075 & 21.300& -22.77& 0.14 & 0.13& Crighton+15\\ 
&&21.300 & 21.800& -23.77& 0.30 & 0.18& Crighton+15\\ 
\tlya & 4.05 &11.0 & 22.0&0.91&0.09 & 0.09& Becker+13\\ 
\tlya & 4.15 &11.0 & 22.0&0.98&0.10 & 0.10& Becker+13\\ 
\tlya & 4.25 &11.0 & 22.0&1.02&0.10 & 0.10& Becker+13\\ 
\tlya & 4.35 &11.0 & 22.0&1.07&0.11 & 0.11& Becker+13\\ 
\tlya & 4.45 &11.0 & 22.0&1.13&0.11 & 0.11& Becker+13\\ 
\tlya & 4.55 &11.0 & 22.0&1.20&0.12 & 0.12& Becker+13\\ 
\tlya & 4.65 &11.0 & 22.0&1.24&0.12 & 0.12& Becker+13\\ 
\tlya & 4.75 &11.0 & 22.0&1.42&0.14 & 0.14& Becker+13\\ 
\tlya & 4.85 &11.0 & 22.0&1.50&0.15 & 0.15& Becker+13\\ 
\lox & 4.08 &17.80 & $\infty$&0.54&0.10 & 0.10& Crighton+18\\ 
\lox & 4.55 &17.80 & $\infty$&0.52&0.11 & 0.11& Crighton+18\\ 
\lox & 5.05 &17.80 & $\infty$&0.67&0.12 & 0.12& Crighton+18\\ 
\hline 
\end{tabular} 
\end{table*}

\section{Constraints on the high redshift column density distribution}
\label{sec:fN}

For a combination of scientific, pragmatic, and historical reasons,
we are motivated to utilize the LLS \loz\ results and previous measurements
from the GGG survey (and literature) to derive a first estimate for the \NHI\ frequency
distribution \fnhi\ at $z \sim 5$.  Scientifically, changes 
in the shape or normalization  of \fnhi\ with redshift 
may offer physical insight
into evolution in the IGM, CGM, and ISM of high-$z$ galaxies.
This holds despite the fact that we have no physical model for \fnhi\
and only initial efforts to estimate it within cosmological simulations
\citep[e.g.][]{McQuinn11, rahmati+13, 2015MNRAS.452.2034R, 2018A&A...614A..31B, 2018MNRAS.474.3032R, 2018MNRAS.476.3716R}.
%
%
Our approach to evaluating \fnhi\ follows the methodology developed
in \citetalias{pmo+14}.  The mathematical model adopted is a 
Piecewise Cubic Hermite Interpolating Polynomial 
evaluated here with the {\sc PchipInterpolator} algorithm in 
{\sc scipy} (v1.0.1)\footnote{We caution that this algorithm in {\sc scipy}
has changed
at least once in the past two years and therefore the results given here
may depend on the precise version of {\sc scipy}.  We will 
endeavor to update our model in the {\sc pyigm} package
as {\sc scpiy} evolves.}.
The model is parameterized by seven ``pivots" at specific $\log \mNHI$ values.
As in \citetalias{pmo+14}, we assume 
$(1+z)^\gamma$ redshift evolution in the
normalization of \fnhi, and fix $\gamma = 1.5$ for this first model
and pivot at $z_{\rm pivot} = 4.4$. 
The number and locations of the pivots 
correspond to \NHI\ values where the model is
likely to be well-constrained by the data (e.g.\ at 
$\mNHI \approx 10^{17}$\cmm\ from LLSs) 
with additional, ``outer" pivots at low and high \NHI\ to bound the model.
After some experimentation, the pivot positions were slightly
adjusted to yield better results. 

Regarding the observational constraints, we included the following
set of measurements, as summarized in Table~\ref{tab:z5_fn_constraints}:
\begin{itemize}
\item the \tlya\ measurements from \cite{Becker13} for $z>4$;
\item the \lmfp\ measurements from the GGG survey \citep{Worseck_2014};
\item the \ldla\ and \fnhi\ measurements of DLAs from the GGG survey
\citep{Crighton15};
\item the \llls\ measurements from this paper.
\end{itemize}
Evaluations of \fnhi\ to compare against the other, integral constraints
were performed with the {\sc pyigm} 
package.

We performed a Monte Carlo Markov Chain (MCMC) analysis with 
the {\sc pymc3} (v3.4.1)  software package
and its ``slice sampling" method.
We adopted the reported uncertainties for each measurement and 
assumed a Gaussian distribution in the deviate evaluation.
In addition, our final results were derived assuming a Gaussian
prior on each pivot within a standard deviation of $\pm 0.5$\,dex centered
on estimates from trial runs.  
We constructed a set of 8
chains, each with 15000 links modulo a 1,000 step burn-in phase.
The collated set of 112,000 links yields the probability density functions describing the
\fnhi\ function.  

\begin{figure*}
\begin{center}
\includegraphics[width=0.7\textwidth]{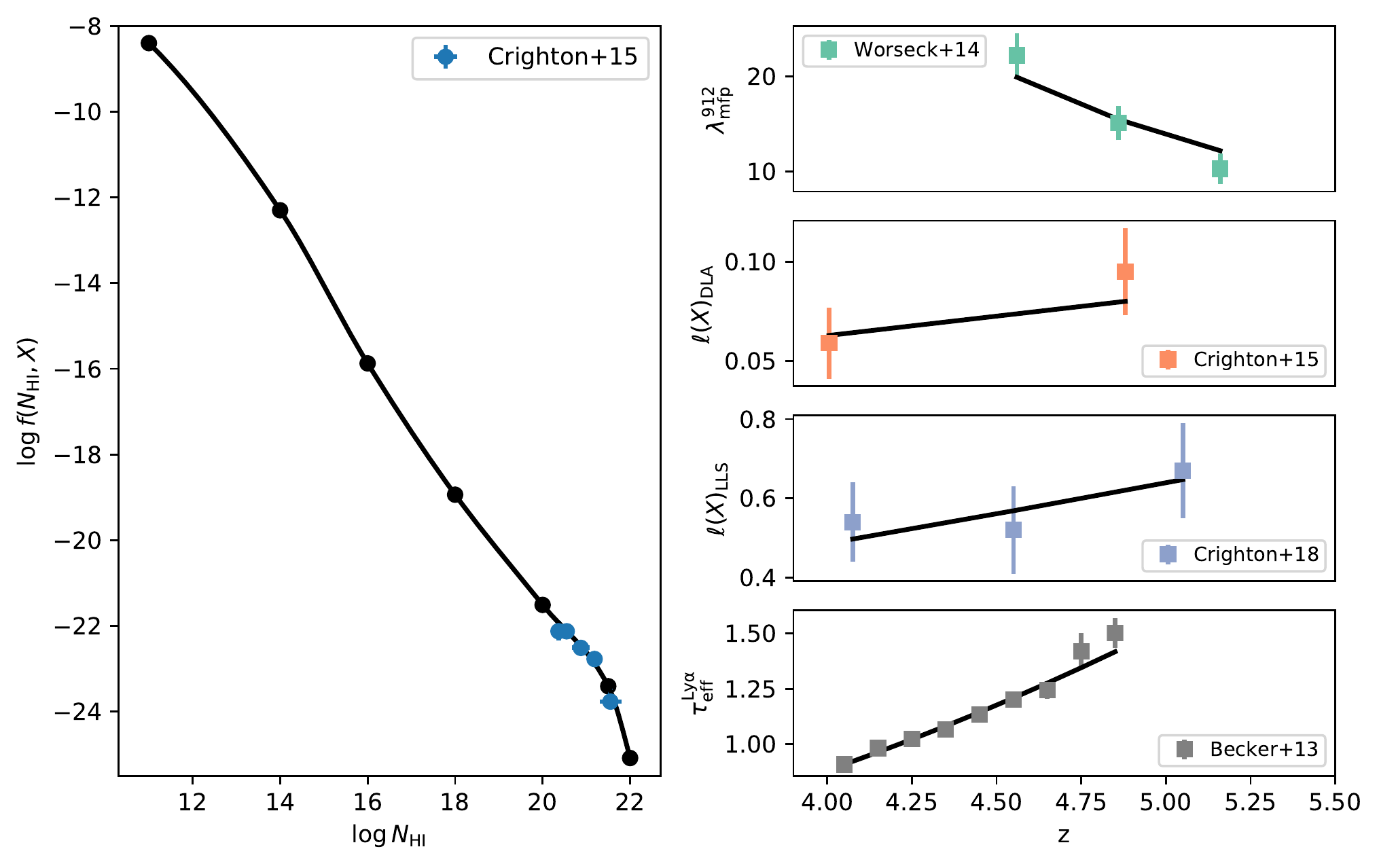}
\caption{{\label{f:fN_z5}
Best-fitting model of \fnhi\ at $z=4.4$. (left) The dots and black curve show the best fit model
to the estimations of \fnhi\ at $\mNHI > 10^{20}$\cmm\ 
from the GGG survey \citep[blue points][]{Crighton15}.
(right) Integral constraints on \fnhi\ from the GGG
survey, including this work, and for the \lya\ forest
\tlya\ measurements from \citep{Becker13}.
The model is an excellent description of the data
except for the highest redshift \tlya\ measurements.
}}
\end{center}
\end{figure*}

\begin{figure*}
\begin{center}
\includegraphics[width=0.7\textwidth]{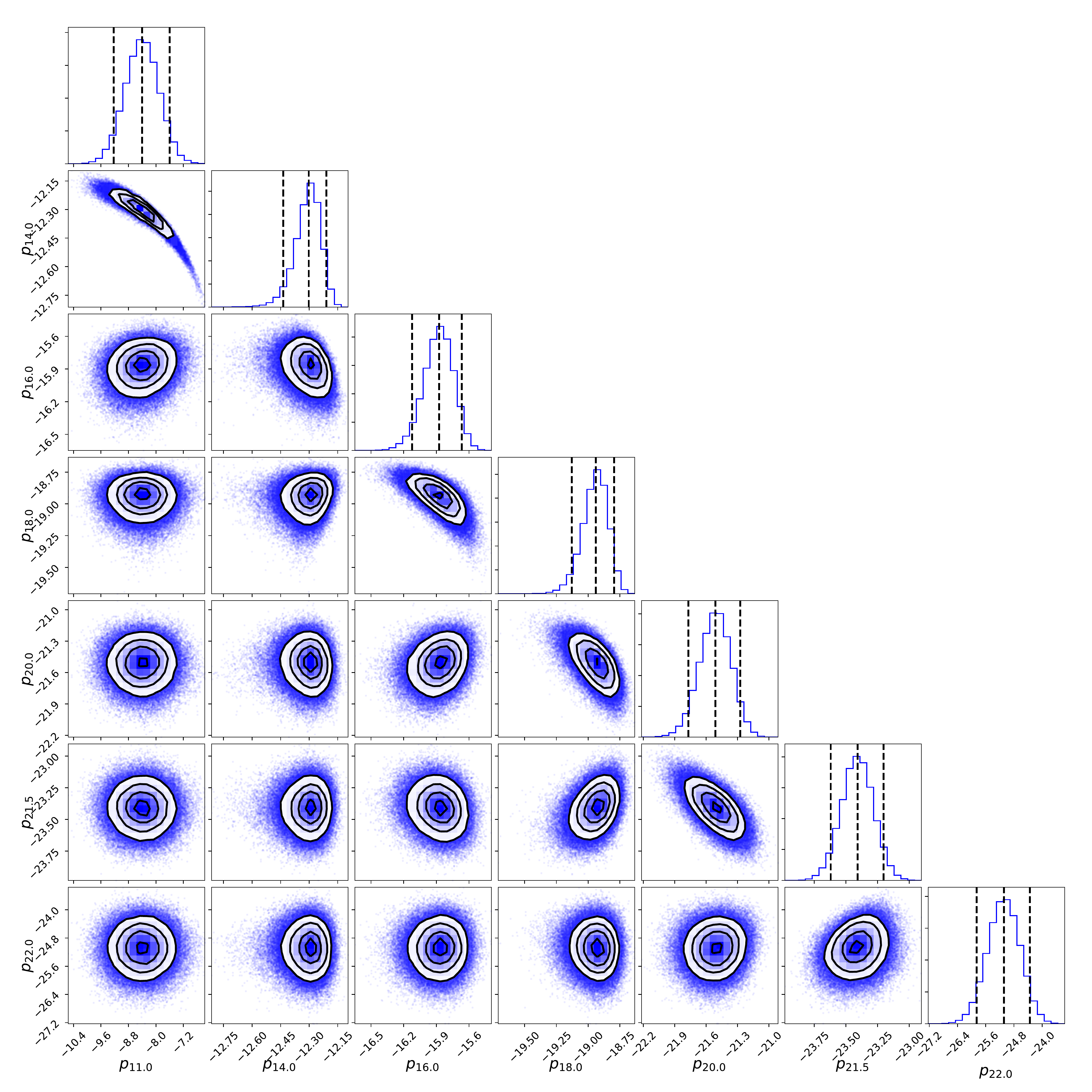}
\caption{{\label{f:corner_z5}
Corner plot analysis of the MCMC results for the $z \sim 5$
analysis (see Fig.\ \ref{f:fN_z5}).  Note that the outer
pivots ($p_{11.0}, p_{22.0}$) 
are the most poorly constrained. Note also also the significant
degeneracy between neighbouring pivots, as expected.
These issues aside, the inner pivots are well constrained.
}}
\end{center}
\end{figure*}

Figure \ref{f:fN_z5} presents the primary results, with the left panel showing the \fnhi\ function evaluated at $z=4.4$ and the right panels comparing the integral constraints against this model.
The model provides an excellent description of the integral constraints
on \fnhi\ and the estimations of \fnhi\ in the DLA regime.
Uncertainties in the model have been derived from the MCMC chain
and are described by the ``corner" plots in Figure~\ref{f:corner_z5}.
There is significant correlation between neighbouring points in the
model, but the inner pivots are otherwise well-behaved.
As should be expected, uncertainties in the outer pivots, 
used primarily to bound \fnhi, are the largest.
The results are summarized in Table \ref{tab:fn_results}
which lists the median normalization of 
each pivot in the \fnhi\ model, at $z=4.4$, and the
68\%\ confidence interval from the probability density functions. 

\begin{table}
\begin{center}
\caption{\fnhi\ modelling results. The prior and resulting median value for each parameter in the model is listed, together with the 16th and 84th percentile deviations from the median in the parameter probability distribution, indicating the 68\% confidence interval.\label{tab:fn_results}}
\begin{tabular}{ccccc}
\hline 
Parameter & Prior$^a$ & Median & 16th       & 84th\\ 
          &           &        & percentile & percentile\\ 
\hline 
\multicolumn{5}{|c|}{Results for $z \sim 5$} \\ 
$p_{11.0}$& $G(-8.38, 0.5)$& $-8.40$& $0.50$& $0.49$\\ 
$p_{14.0}$& $G(-12.30, 0.5)$& $-12.30$& $0.07$& $0.06$\\ 
$p_{16.0}$& $G(-16.00, 0.5)$& $-15.87$& $0.14$& $0.13$\\ 
$p_{18.0}$& $G(-19.12, 0.5)$& $-18.94$& $0.11$& $0.09$\\ 
$p_{20.0}$& $G(-21.20, 0.5)$& $-21.51$& $0.16$& $0.15$\\ 
$p_{21.5}$& $G(-23.60, 0.5)$& $-23.41$& $0.13$& $0.12$\\ 
$p_{22.0}$& $G(-24.94, 0.5)$& $-25.08$& $0.46$& $0.45$\\ 
\multicolumn{5}{|c|}{Results for $z \sim 2-5$} \\ 
$\gamma$& $G(1.5, 0.5)$& $1.77$& $0.06$& $0.06$\\ 
$p_{11.0}$& $G(-8.38, 0.5)$& $-8.00$& $0.10$& $0.10$\\ 
$p_{14.0}$& $G(-12.30, 0.5)$& $-12.41$& $0.02$& $0.02$\\ 
$p_{16.0}$& $G(-16.00, 0.5)$& $-15.79$& $0.04$& $0.04$\\ 
$p_{18.0}$& $G(-19.12, 0.5)$& $-19.20$& $0.05$& $0.04$\\ 
$p_{20.0}$& $G(-21.20, 0.5)$& $-20.97$& $0.02$& $0.02$\\ 
$p_{21.5}$& $G(-23.60, 0.5)$& $-23.56$& $0.03$& $0.03$\\ 
$p_{22.0}$& $G(-24.94, 0.5)$& $-25.16$& $0.12$& $0.12$\\ 
\hline 
\end{tabular} 
\end{center}
{$^a$}$G(\mu,\sigma)$ indicates a Gaussian distribution with mean $\mu$ and standard deviation $\sigma$.\\ 
\end{table}

\begin{figure*}
\begin{center}
\includegraphics[width=0.7\textwidth]{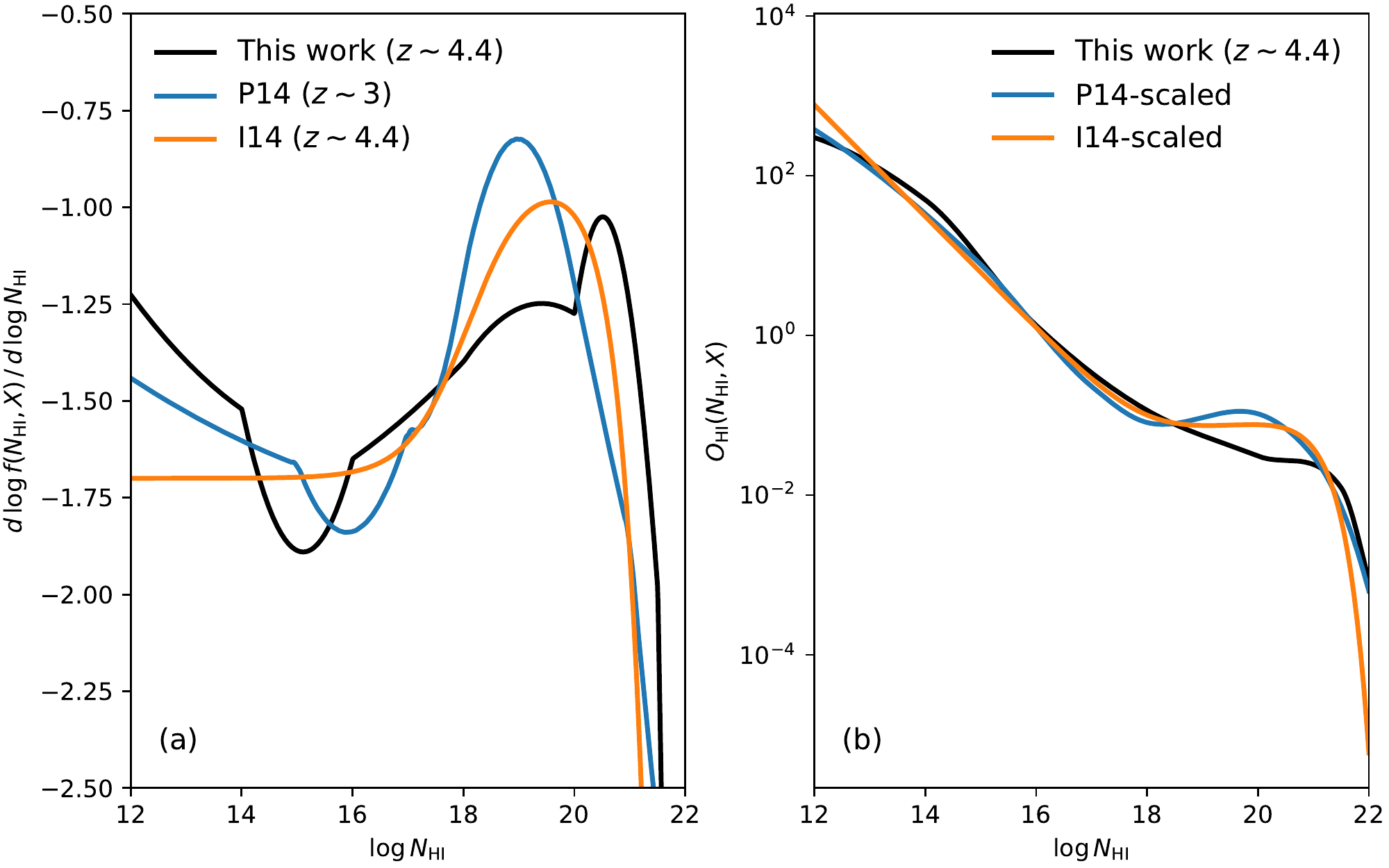}
\caption{\label{f:dlogf}
(a) Plot of \dlogf\ for the \fnhi\ model fit to the
$z \sim 5$ constraints (black) and the corresponding curves
for the \fnhi\ models of \citetalias{pmo+14} and 
\citet{inoue+14}.  
All \fnhi\ models show a steepening of \fnhi\ in the
\lya\ forest regime ($\mNHI \sim 10^{\text{12--16}}$\cmm)
and then a flattening in optically thick gas.
The model also steepens through the DLA regime.
(b) Plot of $\mON \equiv \mfnhi \, \mNHI$ for the same
models with the \citetalias{pmo+14} and \citet{inoue+14}
models extrapolated to $z=4.4$ with
a $(1+z)^{1.5}$ power-law.  
The greater flattening of 
the \citetalias{pmo+14} and \citet{inoue+14} models 
in \fnhi\ manifests 
as an excess in \ON\ 
at $\mNHI \sim 10^{20}$\cmm.
}
\end{center}
\end{figure*}

It is clearly evident from Fig.\ \ref{f:fN_z5} that the \fnhi\ distribution is not a single power-law
over the $\sim$10 orders-of-magnitude in \NHI\ that quasar spectra probe.  This is further illustrated
in Fig.\ \ref{f:dlogf}a where we plot \dlogf\ against $\log \mNHI$.
At $\mNHI \approx 10^{12}$\,\cmm, corresponding to the \lya\ forest
and approximately the mean density of the universe at $z \sim 5$,
$\mfnhi \propto \mNHI^{-1.2}$ and then steepens
as \NHI\ increases to $\approx 10^{16}$\,\cmm.  
The distribution then flattens as one enters
the optically-thick regime ($\mNHI \approx 10^{17}$\,\cmm), 
a behaviour expected in the transition from
ionized to neutral gas \citep[e.g.][]{zheng02}.
Figure~\ref{f:dlogf} also compares these results on \fnhi\ at $z \sim 4.4$
with estimates at lower redshifts, $z\sim2$--3, based on similar constraints and
analysis \citep{pmo+14,inoue+14}. 
The \dlogf\ behaviour is qualitatively similar between
the two models, although the flattening in \fnhi\ is more
pronounced at $z \sim 3$.
This leads to the bump in $\mON \equiv \mfnhi \mNHI$
for $z\sim3$ around $\mNHI \sim\ 10^{20}$\,\cmm, evaluated and shown in Fig.\ \ref{f:dlogf}b.
This bump implies an increase in the relative number of high-column density LLSs and low-column density DLAs from $z\sim$5 to $z \sim 2$--3, which at present appears somewhat contrary to predictions from cosmological 
simulations \citep[e.g.][]{fop11}.

\begin{figure*}
\begin{center}
\includegraphics[width=0.7\textwidth]{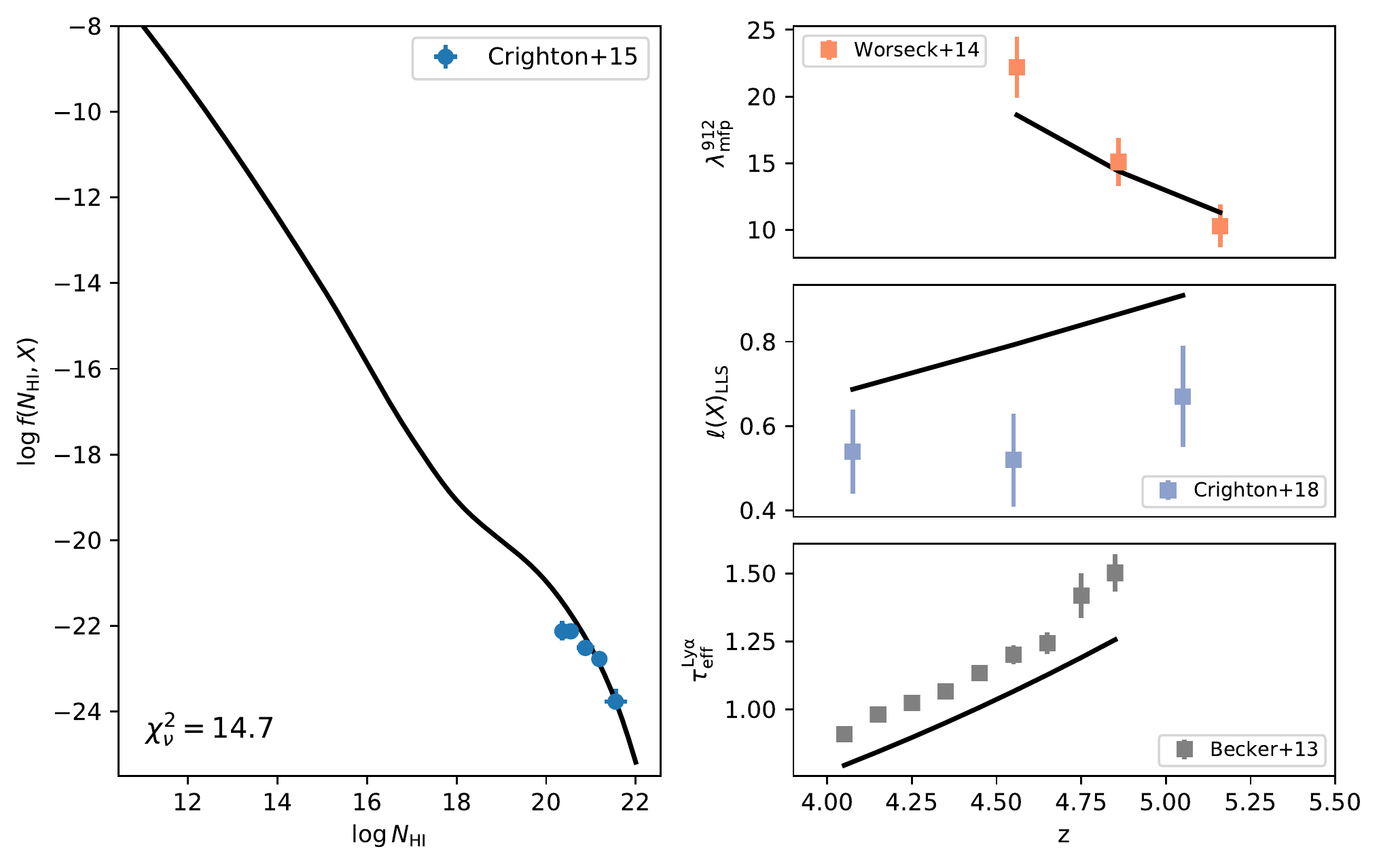}
\caption{\label{f:z3_shape}
Best-fit model of the $z \sim 5$ constraints 
(Table~\ref{tab:z5_fn_constraints}) adopting the shape of \fnhi\ 
from \citetalias{pmo+14} and fitting for the power-law exponent $\gamma$
that describes the redshift evolution (and therefore the
normalization at $z>4$).  We find $\gamma = \bestgamma$
but this model significantly under-predicts the opacity of
the \lya\ forest \tlya\ while over-predicting the incidence
of LLSs and DLAs.
}
\end{center}
\end{figure*}

To further examine redshift evolution in the shape of
\fnhi, we have adopted the form of
\fnhi\ from \citetalias{pmo+14} and derived the most likely value for $\gamma$
by fitting to the $z \sim 5$
constraints in Table~\ref{tab:z5_fn_constraints}.
We derive $\gamma = \bestgamma$ and 
the model is compared to the observational constraints
in Figure~\ref{f:z3_shape}.
This model greatly over-estimates the 
incidence of $\mtauLL \ge 2$ systems (including DLAs)
and under-predicts the opacity of the \lya\ forest.
Because modifying the normalization would improve one at the
expense of the other,  we conclude that the \citetalias{pmo+14} \fnhi\ model
is a poor description of the observations at $z \sim 5$.

\begin{figure*}
\begin{center}
\includegraphics[width=0.7\textwidth]{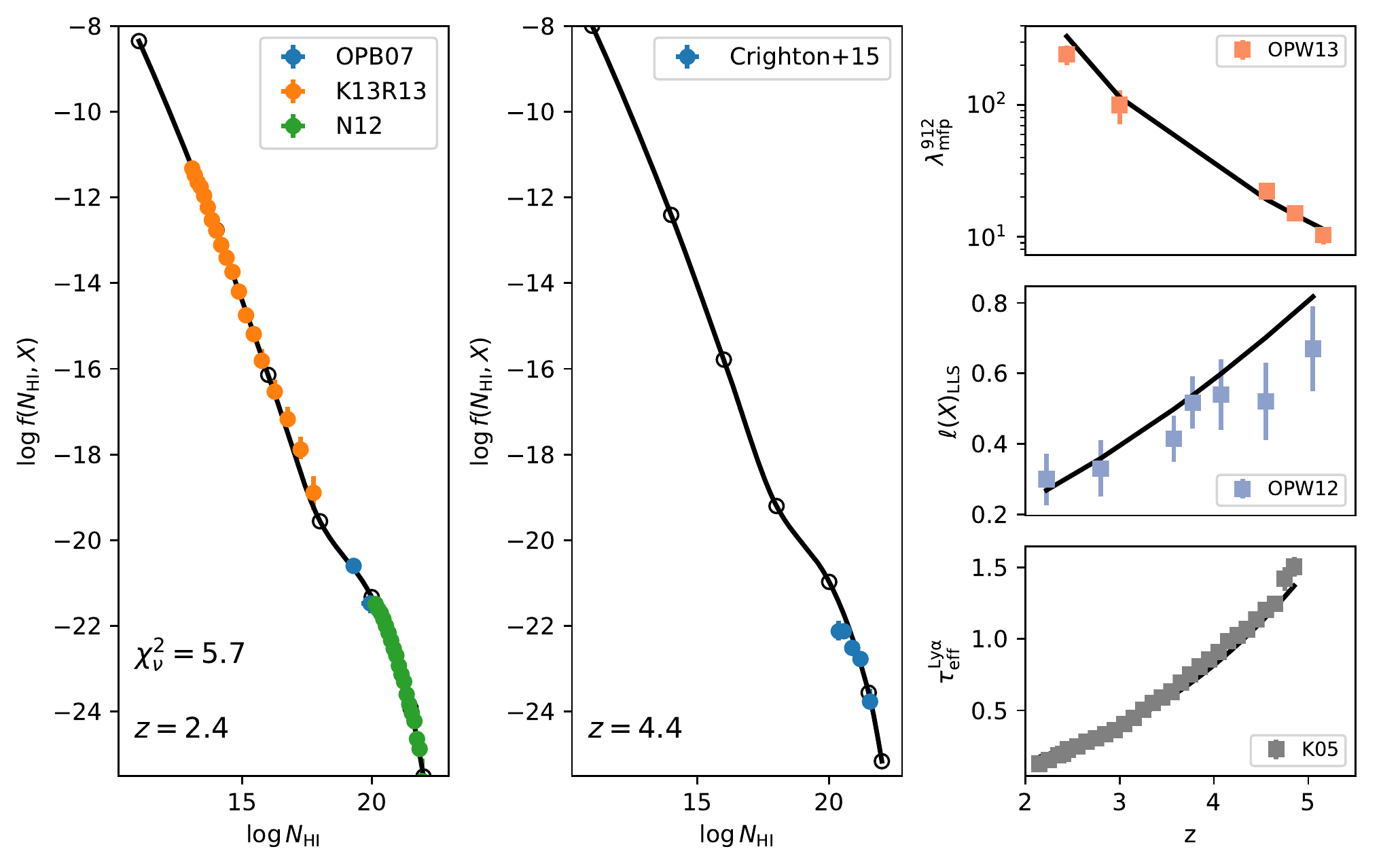}
\caption{{\label{f:enchilada}
Comparison of an 8-parameter \fnhi\ model fit to constraints
from $z \sim 2-5$.  The two left panels show model (black line
and dots at pivots) evaluated at $z=2.4$ and 4.4 compared to
\fnhi\ values from the literature and the GGG survey.
The right-hand panels compare the model to integral constraints
including the mean-free path \citep[\lmfp;][]{OMeara13,Worseck_2014},
the incidence of LLSs \citep[\lox;][this work]{Fumagalli13,Prochaska_2010},
and the effective \lya\ opacity \citep[\tlya;][]{Becker13}.
We report a reduced $\chi^2_\nu = \redchi$ assuming normal
distributions on the uncertainties and formally rule out this
model indicating intrinsic evolution in the shape of \fnhi.
The best-fit exponent for a power-law evolution in the normalization
is $\gamma = \bestgtwo$ implying a significant increase in the
neutral fraction of gas on all scales.
}} 
\end{center}
\end{figure*}

Therefore, we have further generalized
the \fnhi\ model, combining the 7-pivot model above with a
free parameter for $\gamma$. We performed an 
MCMC analysis with all of the $z>3$ constraints of \citetalias{pmo+14}, several 
additional $z<4$ constraints from the literature, and all of
those from Table \ref{tab:z5_fn_constraints}.
Figure \ref{f:enchilada} presents the model compared to the
observational data.  
Qualitatively, the model offers a good approximation to nearly all
of the observational constraints at $z \sim 2$--5.  The clear
exception is the incidence of DLAs at $z>4$ which is somewhat
over-predicted by the model.  
Similarly, this model systematically overestimates
our new \lox\ measurements for $z>4$ LLS.
We also note that the \lox\ model does not cluster LLS
as we have;  a proper treatment would likely exacerbate
the tension between model and data.
Less obvious is the fact that 
the observational constraints on the \tlya\ measurements and the
$z \sim 2$ \fnhi\ values for the \lya\ forest are sufficiently
precise that the resultant $\chi^2$ is large.
We report a reduced $\chi^2_\nu = \redchi$ per degree of freedom in the model,
with 74\%\ of the contribution to this coming from the \tlya\ measurements
of the \lya\ forest data.
We therefore conclude that
the data require evolution in the shape of \fnhi\ over
the few Gyr from $z \approx 5$ to 2.  
A similar conclusion was reached by \cite{inoue+14}
at lower redshifts.
As emphasized in Fig.\ \ref{f:dlogf}, the starkest difference in the \fnhi\ shapes at $z\sim2$ and 5 is the additional flattening in the former at $\mNHI \sim 10^{20}$\,\cmm. 
We encourage further exploration into the CGM and ISM of galaxies
in hydrodynamic simulations to further interpret these results.
We also note that future work should examine \fnhi\ models with
an evolving shape to attempt to fully describe the 
data and understand the reasons for its evolution \citep[see also][]{inoue+14}.

The other important result from our analysis is 
$\gamma = \bestgtwo$ which indicates a substantial
evolution in the normalization of \fnhi.  
A universe where the structures yielding optically thick
gas is non-evolving would yield $\gamma = 0$.
And given that we expect the number density of 
structures (e.g.\ galactic halos) to {\it decrease}
with increasing redshift \citep[e.g.][]{Fumagalli13},
this implies an increase in the neutral fraction of
such gas \citep[see also][]{Worseck_2014}.
Furthermore, the fact 
this holds approximately independent
of \NHI, implies an increase in the
neutral faction of gas on all scales.

\section{Conclusions}

We have presented LLS survey, drawn from the homogeneous GGG sample of \nqso\ $z\sim5$ quasar spectra, extending to the highest resshifts to date, $\mzlls=3.5$--5.4. Using mock spectra to determine the required corrections from the raw LLS counting statistics determined by two different authors, we estimated the incidence of LLSs per unit redshift at $z\approx$4.4 to be $\mloz = \lzfull \pm \slzfull$. Breaking the results into 3 redshift bins, and comparing it with previous results from LLS surveys at $\mzlls<4$, we find that the incidence rate decreases with time consistent with a simple power-law in redshift: $\mloz = \ell_* [(1+z)/4]^\alpha$ with $\ell_* = \lstar$ and $\alpha = \aval$ (68\% confidence intervals). Our new measurements do not bear out hints for a rapid decline in \loz\ from $z\approx$4.2 to 3.6, as hinted at (with relatively large uncertainties) in the LLS survey of SDSS spectra by \citep{Prochaska_2010}. That is, we do not detect any clear post-reionization effects in LLS evolution. The simple physical motivation for the power-law evolution described by \citet{Fumagalli13}, in which LLSs only arise in galaxy haloes larger than a characteristic mass, therefore remains a plausible model.

By combining our measurements of the LLS incidence rate with the DLA incidence rate and distribution function \citep{Crighton15}, and mean-free path measurements \citep{Worseck_2014}, from the same GGG spectra, we are able to constrain the shape of the $z\sim5$ column density distribution function, \fnhi, below our detection threshold for LLSs ($\mNHI < 10^{17.5}$\,\cmm). A 7-pivot cubic Hermite interpolating polynomial self-consistently matches these constraints on \fnhi\ to measure its normalisation and shape over $\sim$10 orders of magnitude, $10^{12} \la \mNHI \la 10^{22}$\,\cmm\ -- see Fig.\ \ref{f:fN_z5}. However, the previously-reported model of \fnhi\ at $z\sim2$--3 cannot consistently match the $z\sim5$ data (allowing for a re-scaling of the normalisation). Nor does a similar model of the combined $z=2$--5 dataset, with an unevolving shape for \fnhi, provide a statistically acceptable fit. We are therefore compelled to conclude that the shape of \fnhi\ evolves from $z\sim5$ down to $z\sim2$--3. The independent models at these redshifts differ most starkly at column densities around $\mNHI \sim 10^{20}$\,\cmm, so we encourage the specific exploration of the CGM and ISM of hydrodynamic galaxy simulations to understand this evolution.

\section*{Acknowledgements}

 NHMC and MTM thank the Australian Research Council for
 \textsl{Discovery Project} grant DP130100568 which supported this
 work. Our analysis made use of \textsc{astropy} \citep{Astropy13} and
 \textsc{matplotlib} \citep{Hunter07}. The authors wish to acknowledge the
 significant cultural role that the summit of Maunakea has always had
 within the indigenous Hawaiian community.  We are most fortunate to
 have the opportunity to conduct observations from this mountain.

\appendix

\section{Mock spectra from IGM simulations}
\label{a:mocks}

\subsection{Creating the mock spectra}

To quantify how well our search procedure can find Lyman limit systems in the real spectra, we generate a set of mock spectra with a known distribution of Lyman limit systems. To create these mocks we require simulated absorption from the IGM. In our previous DLA analysis \citep{Crighton15} we generated IGM absorption using a distribution of Voigt profiles, with a simple line clustering model to match the observed \Lya\ forest. In this work we instead use numerical simulations to produce IGM absorption, which we expect will provide a better approximation to line clustering in the real Universe. The details of the simulations are provided in Section \ref{a:sims}.

\begin{figure}
\begin{center}
\includegraphics[width=0.84\columnwidth]{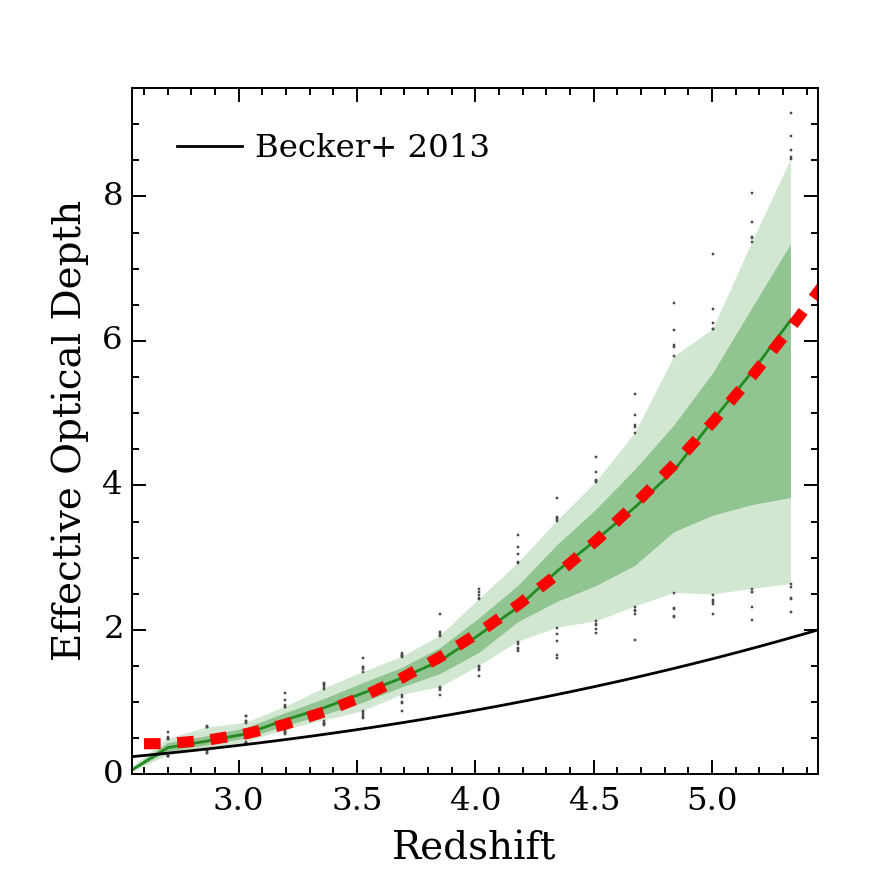}
\caption{{\label{f:taulya}
The effective \Lya\ optical depth as a function of redshift. The green shaded regions shows the range in the simulated skewers (the median, 68\%, and 95\% percentiles along with outliers) and the black line shows the observed values from \citet{Becker13}. The red line shows the second order polynomial fit to the median optical depth, which we used to correct the simulated optical depths to match the \citeauthor{Becker13} relation.
}}
\end{center}
\end{figure}

We create one simulated `skewer' to match each GGG quasar sightline. The optical depth due to \Lya, $\tau_{\mLya}$, is calculated along this skewer from just above the largest quasar redshift in the GGG sample, $z=5.5$, down to $z=2.8$, which corresponds to the blue cutoff of the GMOS spectra at 4620\,\AA. We measure the median effective optical depth $\tau_{\rm eff} \equiv -\ln\langle T \rangle$ as a function of redshift for the ensemble of skewers, where $T$ is the fraction of quasar flux transmitted, given by $\exp(-\tau_{\mLya})$.
We then compare this to the observed optical depth from \citet{Becker13}.  The result is shown in Fig.\ \ref{f:taulya}. The effective optical depth for the mocks is about twice as large as the observed value at $z=4$, and this discrepancy increases with redshift. Therefore, we apply a redshift-dependent scaling to the optical depth in the simulated skewers such that the median $\tau_{\rm eff}$ after scaling matches the Becker et al. relationship. We apply this scaling by fitting a second order polynomial to the ensemble median effective $\tau_{\mLya}$, and then multiplying $\tau_{\mLya}$ by the \citet{Becker13} relation and dividing by the polynomial fit.

Once we have corrected $\tau_{\mLya}$, we introduce absorption from the higher order Lyman series using wavelengths and oscillator strengths from \citet{Morton03}. To identify Lyman limit systems and measure an approximate column density distribution, we identify peaks in the $\tau_{\mLya}$ distribution and convert these to a column density using the relation from \citet{Draine11}, assuming $b=20$ \kms. This approximation will be relatively accurate for lines which have a much higher optical depth than the median value, but is likely to break down for low column density systems which are strongly blended (i.e. $\log \mNHI \lesssim 10^{14}$). We then add Lyman continuum absorption for all systems with a Lyman limit optical depth $>$0.01. By visual inspection of the mock spectra, we found that the resulting transmission spectra contained too little absorption at wavelengths shorter than the quasar Lyman limit, which implies that there are too few strong absorbers and LLSs ($\mNHI \gtrsim 10^{15}$ \cmm). These attenuate a significant amount of flux at such wavelengths. Because the simulations do not include radiative transfer calculations or use a self-shielding prescription, this lack of higher column density absorbers is unsurprising: high density clouds will be artificially highly ionised in the simulations.

\begin{figure}
\begin{center}
\includegraphics[width=0.84\columnwidth]{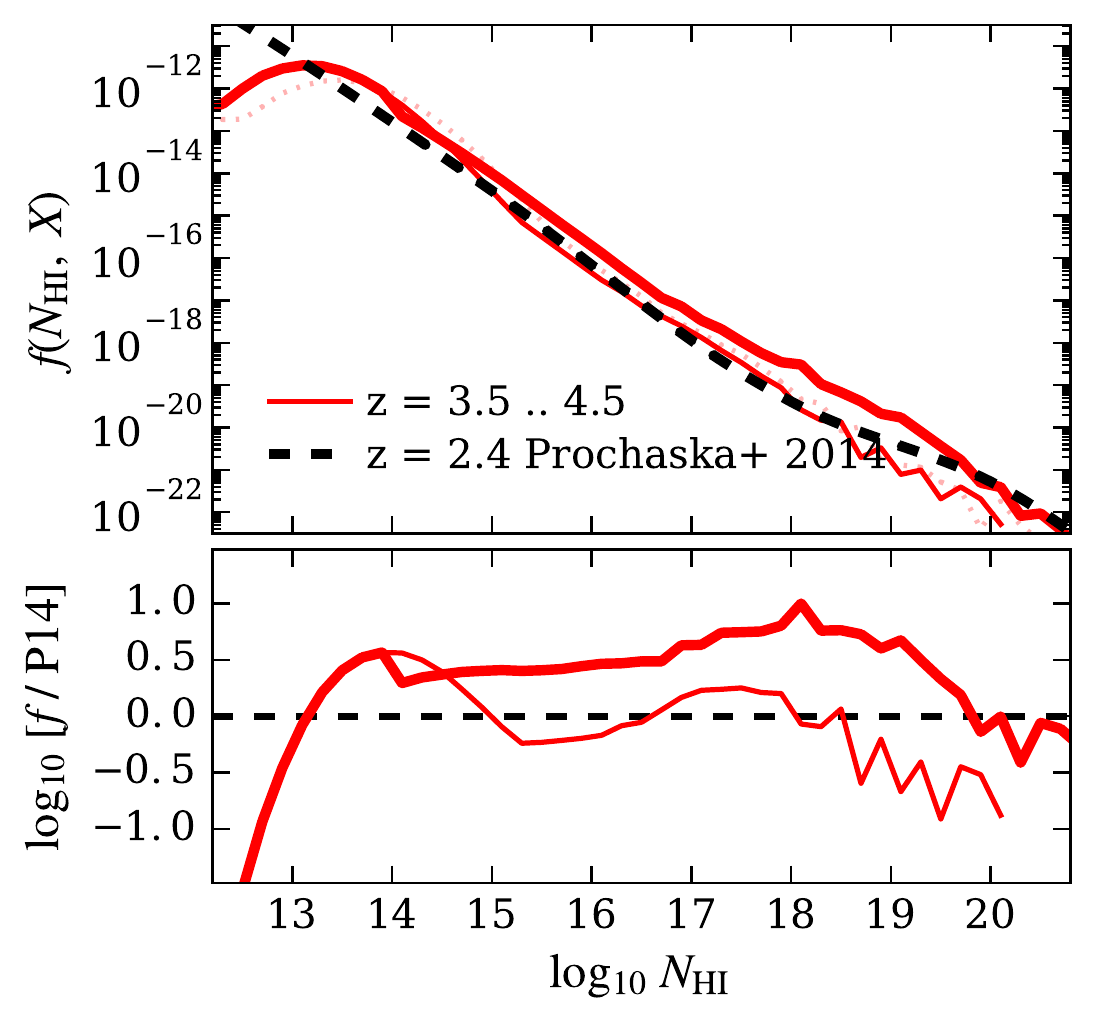}
\caption{{\label{f:cddf}
The column density distribution function generated from the simulated IGM optical depths. The thin and thick solid lines show the distributions before and after correction, respectively, to match the shape (but not normalisation) of the $z=2.4$ distribution from \citet{pmo+14} (dashed line).%
}}
\end{center}
\end{figure}

\begin{figure}
\begin{center}
\includegraphics[width=0.84\columnwidth]{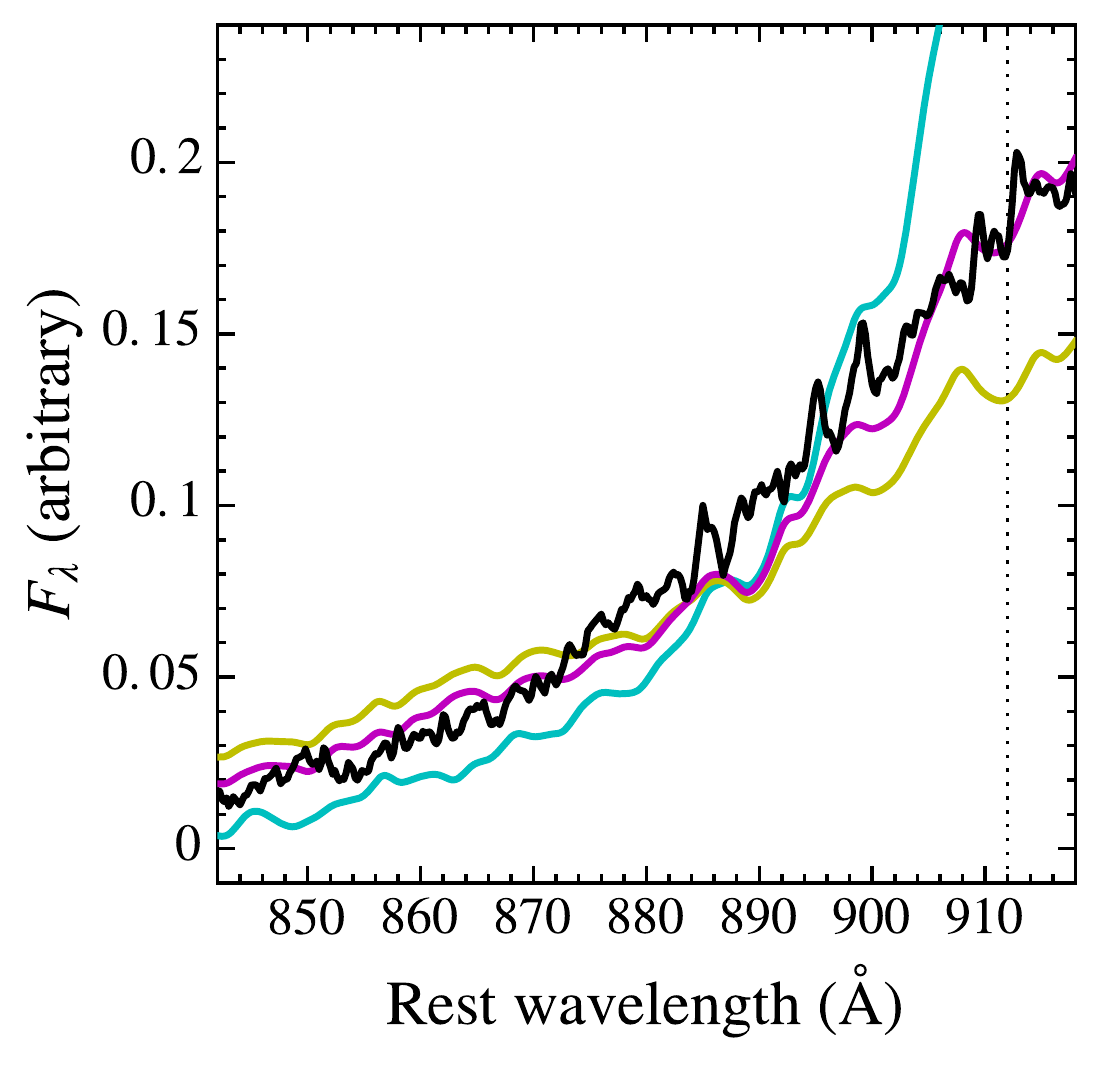}
\caption{{\label{f:LLstack}
An average of all the GMOS spectra, compared to mock spectra with different \Lya\ forest column density distributions. The thick black line shows the average of the real GMOS spectra, plotted against wavelength in the quasar frame. The vertical dashed line shows the quasar Lyman limit. Coloured lines show the averages from mocks generated using $\log \mNHI$ offsets at \lNHI values $[15, 16, 18]$ of $[+0.5, +2, +2]$ (yellow), $[+1, +2, +2]$ (magenta) and $[+1.5, +2 ,+2]$ (cyan). The mock averages are scaled to match the real average in the range $800$ \AA\ $< \lambda_{\rm rest} < 912$\,\AA, and smoothed for clarity. The shape of the mean spectrum is quite sensitive to the frequency of systems at $15 \lesssim \lNHI \lesssim16$. We adopt final offsets of $[+1, +2.2, +2]$, which provide a good match to the observed Lyman limit average.
}}
\end{center}
\end{figure}

We test this idea with Fig.\ \ref{f:cddf} where the thin solid (red) line shows the column density distribution per unit absorption distance in the simulations over the range $3.5<z<4.5$ compared to the observed distribution at $z\sim 2.4$ from \citet{pmo+14}. We assume that there is little evolution in the distribution shape between $z=2.4$ to $z=4$ apart from a change in the normalization corresponding to the $\tau_\text{eff}$ evolution. In this case there does indeed appear to be a deficit of systems with $N=10^{\text{15--16}}\,$\cmm\ and $N>10^{18}\,$\cmm. To correct this deficit, we multiply the column density of $\tau$ peaks by an amount which changes with column density, i.e.\ we add an offset of (0, 1, 2.2, 2) to $\log \mNHI$ for each system at \lNHI\ reference values of (14, 15, 16, 18) and a linearly interpolated offset for systems with other $\log \mNHI$ values. These reference $\log \mNHI$ values were chosen arbitrarily, to provide enough flexibility to adjust the column densities in the range $15< \lNHI < 16$ to better match the observed column density distribution. The offsets at these reference values were found by generating many sets of mock spectra in a parameter grid of offset values, and choosing the set of offset parameters such that a rest-frame average of the mock spectra matched a rest frame average of the real spectra in the Lyman limit region ($\lambda_{\rm rest}= 800$--$912$\AA, see Fig.~\ref{f:LLstack}). Applying this correction produces the column density distribution shown by the thick solid (red) line in Fig.\ \ref{f:cddf}. This correction flattens out the `dips' in the high \NHI\ ranges such that it better matches the shape of the observed distribution.

\subsection{Simulations used for the mock spectra}
\label{a:sims}

The simulations were run using the open-source, cosmological adaptive mesh refinement and N-body code,
{\sc enzo}\footnote{\urlstyle{rm}\url{http://enzo-project.org}}, which is fully
described in \citet{Bryan_2014}.  To solve the hydrodynamics,
we use the Piecewise Parabolic Method of \citet{Woodward_1984}.
To compute the radiative heating and cooling of the gas, we use the
{\sc grackle}\footnote{\urlstyle{rm}\url{https://grackle.readthedocs.org}} chemistry and
cooling library \citep[version 2.1,][]{Bryan_2014,Kim_2013,Smith17}.  We use the primordial chemistry solver in the {\sc grackle} library to
calculate the non-equilibrium evolution of \HI, \HII, \HeI, \HeII, \HeIII,
and electrons.  The heating and cooling from metals is included by
interpolating from multi-dimensional tables created with the
photo-ionization code, {\sc cloudy} \citep{Ferland_2013}.
For both the primordial elements and metals, we include
photo-ionisation and photo-heating from the UV background model of
\citet{Haardt12}.  We include the effects of star formation
and feedback following the model of \citet{DallaVecchia12},
whereby thermal energy is injected in the grid cell hosting a star
particle to a level high enough to make the sound crossing time
shorter than the cooling time, a numerically-based criterion that
\citet{DallaVecchia12} have shown to be successful at creating
realistic outflows.

The simulation consists of a 100 Mpc/h box with 256$^{3}$ dark matter particles and grid cells on the root grid.  We zoom into a region 25 Mpc/h on a side with two nested refinement levels, each refining by a
factor of 2 in length, or 8 in mass.  The dark matter particle mass
within the high-resolution region is roughly 3$\times10^{7}$\,M$_{\odot}$.  We allow 5 additional levels of adaptive mesh refinement
for a maximum spatial resolution of $\sim$2\,kpc/h.  The
high-resolution region was chosen such that it represents an average
region within the larger 100 Mpc/h box.  The simulation and all
associated methods will be described in full detail in Smith \&
Khochfar (in preparation).

The spectra are generated using the
{\sc yt}\footnote{\urlstyle{rm}\url{http://yt-project.org}} simulation analysis
toolkit \citep{Turk11}.  Each spectrum consists of a line
of sight created in the method described by
\citet{Smith11} and updated by \citet{2017ApJ...847...59H}, in which multiple outputs from the
%
%
simulation are used to span the redshift range from z = 5.5 to 3.  At each redshift interval, each line of
sight is randomly oriented to minimize sampling the same structures
and only the high resolution region is used.  For each grid cell
intersected by the line of sight, we extract the \HI\ number density,
temperature, velocity, and redshift, and insert an appropriately
shifted Voigt profile in a synthetic spectrum.

\subsection{Comparison between the mock and observed LLS distribution}

Our aim is for the LLS distribution in the mock spectra to match the distribution of the real LLSs as closely as possible. If this is the case, we can be more confident that the correction factors we derive from the observed to the true \lox\ in the mocks will also be applicable to the real GMOS spectra.

\begin{figure*}
\begin{center}
\includegraphics[width=0.85\textwidth]{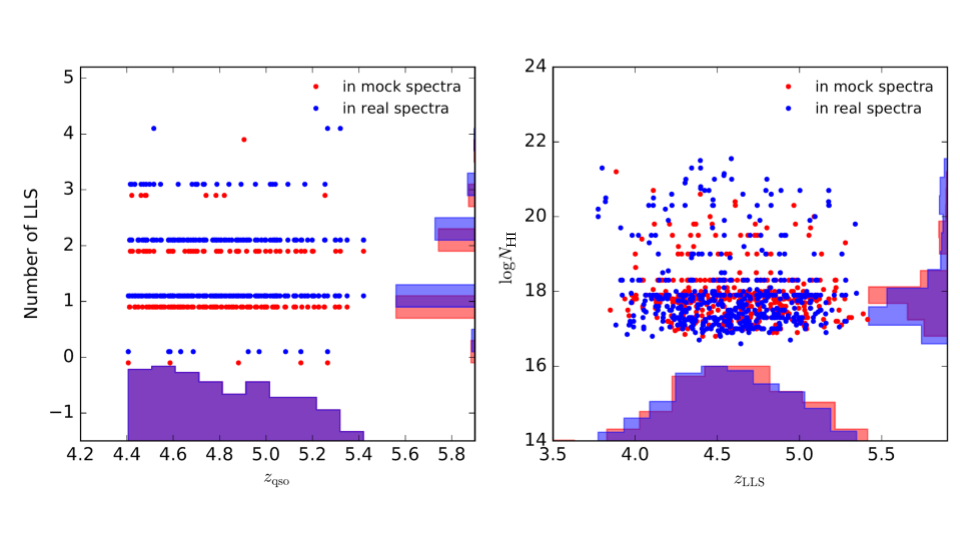}
\caption{{\label{f:cf_mock_real} Left panel: The number of LLSs per sightline identified in the mock and real spectra as a function of the quasar emission redshift. Note that these are LLSs that we identified from the mock spectra, as described in section \ref{s:lls_id}, not the input list of LLSs used in creating the mocks.  The mock and real spectra show a similar distribution (histogram). Right panel: The redshift and \NHI\ distributions for LLSs in the real and mock spectra.}}
\end{center}
\end{figure*}

In Fig.\ \ref{f:cf_mock_real} we compare the redshift distribution, the number of LLSs per sightline, and the column density distribution of the identified LLSs for in the mock sample and real spectra. The redshift distribution for LLSs in the real and mock spectra are very similar. The number of LLSs per sightline are also similar, although there tend to be slightly fewer mock sightlines with 2 or more LLSs compared to the real sightlines. The \NHI\ distribution is similar, although somewhat more partial LLSs ($\mtauLL \lesssim 1$) are identified in the mock spectra compared to those found in the real spectra. Through these checks, together with a visual comparison of each real and mock spectrum, we conclude that the LLS distribution in the mock spectra match the distribution observed in the real spectra closely enough to derive a reliable correction factor for \lox.

\bibliographystyle{mnras}
\bibliography{ggg_lls.bib}

\end{document}